 \documentclass[onecolumn, 12pt, draftclsnofoot, comsoc]{IEEEtran}
 \usepackage{lipsum}
\usepackage{setspace}

\renewcommand{\baselinestretch}{2} 

\usepackage{gensymb}
\usepackage{cite}
\usepackage{amsmath,amssymb,amsfonts}
\usepackage{graphicx}
\usepackage{xcolor}
\usepackage{kotex}  
\usepackage{caption}
\usepackage{subcaption}
\usepackage{caption}
\usepackage{subcaption}

\usepackage{blindtext}
\usepackage{titlesec}

\usepackage{booktabs}
\usepackage{longtable}

\usepackage{stfloats}
\usepackage{float}
\usepackage{wrapfig}
\usepackage{xcolor}
\usepackage{colortbl} 
\usepackage{color,soul}
\usepackage{url}
\usepackage{verbatim}
\usepackage{graphicx}
\usepackage{cite}

\usepackage[acronym,toc,shortcuts]{glossaries}
\newacronym{BS}{BS}{Base Station}
\newacronym{NTN}{NTN}{Non-Terrestrial Networks}
\newacronym{TN}{TN}{Terrestrial Network}
\newacronym{NGMA}{NGMA}{Next-Generation Multiple Access}
\newacronym{NOMA}{NOMA}{Non-Orthogonal Multiple Access}
\newacronym{PD-NOMA}{PD-NOMA}{Power Domain Non-Orthogonal Multiple Access}
\newacronym{CD-NOMA}{CD-NOMA}{Code Domain Non-Orthogonal Multiple Access}
\newacronym{SIC}{SIC}{Successive Interference Cancellation}
\newacronym{UEs}{UEs}{User Equipment's}
\newacronym{UE}{UE}{User Equipment}
\newacronym{UAV}{UAV}{Uncrewed Aerial Vehicles}
\newacronym{SDG}{SDG}{Sustainable Development Goals}
\newacronym{UN}{UN}{United Nations}
\newacronym{AI}{AI}{Artificial Intelligence}
\newacronym{EE}{EE}{Energy Efficiency}
\newacronym{3GPP}{3GPP}{3$^{\text{rd}}$ Generation Partnership Project}
\newacronym{SA1}{SA1}{Service and System Aspects 1}
\newacronym{WG}{WG}{Working Group}
\newacronym{WGs}{WGs}{Working Groups}
\newacronym{PLMN}{PLMN}{Public Land Mobile Networks}
\newacronym{MNOs}{MNOs}{Mobile Network Operators}
\newacronym{LEO}{LEO}{Low Earth Orbit}
\newacronym{KPI}{KPI}{Key Performance Indicators}
\newacronym{GEO}{GEO}{Geostationary Equatorial Orbit}
\newacronym{MEO}{MEO}{Medium-Earth Orbit}
\newacronym{CapEx}{CapEx}{Capital Expenditures}
\newacronym{IMS}{IMS}{IP Multimedia Subsystem}
\newacronym{SA}{SA}{Service \& System Aspects}
\newacronym{CT}{CT}{Core Network \& Terminals}
\newacronym{RIS}{RIS}{Reconfigurable Intelligent Surfaces}
\newacronym{RAN}{RAN}{Radio Access Network}
\newacronym{GNSS}{GNSS}{Global Navigation Satellite System}
\newacronym{NG-RAN}{NG-RAN}{Next Generation Radio Access Network}
\newacronym{O-RAN}{O-RAN}{Open-Radio Access Network}
\newacronym{OMA}{OMA}{Orthogonal Multiple Access}
\newacronym{FDMA}{FDMA}{Frequency Division Multiple Access}
\newacronym{TDMA}{TDMA}{Time Division Multiple Access}
\newacronym{SATCOM}{SATCOM}{Satellite Communications}
\newacronym{ISTN}{ISTN}{Integrated Satellite-Terrestrial Networks}
\newacronym{NGSO}{NGSO}{Non-Geostationary Satellite Orbit}
\newacronym{UPF}{UPF}{User Plane Function}
\newacronym{ISL}{ISL}{Inter-Satellite Link}
\newacronym{QoS}{QoS}{Quality of Service}
\newacronym{LAN}{LAN}{Local Area Network}
\newacronym{WLAN}{WLAN}{Wide Local Area Network}
\newacronym{ISAC}{ISAC}{Integrated Sensing and Communication}
\newacronym{SNO}{SNO}{Satellite Network Operators}
\newacronym{D2M}{D2M}{Direct-to-Mobile}
\newacronym{SMS}{SMS}{Short Messaging Service}
\newacronym{TS}{TS}{Technical Specification}
\newacronym{NR}{NR}{New Radio}
\newacronym{eMBB}{eMBB}{Enhanced Mobile Broadband}
\newacronym{mMTC}{mMTC}{Massive Machine Type Communication}
\newacronym{RF}{RF}{Radio Frequency}
\newacronym{IoT}{IoT}{Internet of Things}
\newacronym{LTE}{LTE}{Long Term Evolution}
\newacronym{NB-IoT}{NB-IoT}{Narrow Band-IoT}
\newacronym{VR}{VR}{Virtual Reality}
\newacronym{AR}{AR}{Augmented Reality}
\newacronym{CSI}{CSI}{Channel State Information}
\newacronym{CSIT}{CSIT}{Channel State Information at the Transmitter}
\newacronym{CSIR}{CSIR}{Channel State Information at the Receiver}
\newacronym{SDMA}{SDMA}{Spatial Division Multiple Access}
\newacronym{RSMA}{RSMA}{Rate-Spitting Multiple Access}
\newacronym{MIMO}{MIMO}{Multiple-Input Multiple-Output}
\newacronym{DoF}{DoF}{Degree of Freedom}
\newacronym{NOUM}{NOUM}{Non-Orthogonal Unicast and Multicast}
\newacronym{uRLLC}{uRLLC}{ultra-Reliable Low-Latency Communications}
\newacronym{LoS}{LoS}{Line-of-Sight}
\newacronym{DL}{DL}{Deep Learning}
\newacronym{RL}{RL}{Reinforcement Learning}
\newacronym{ML}{ML}{Machine Learning}
\newacronym{RRM}{RRM}{Radio Resource Management}
\newacronym{EIRP}{EIRP}{Effective Isotropic Radiated Power}
\newacronym{VSAT}{VSAT}{Very Small Aperture Terminal}
\newacronym{RU}{RU}{Radio Unit}
\newacronym{DU}{DU}{Distributed Unit}
\newacronym{SUs}{SUs}{Satellite Users}
\newacronym{CUs}{CUs}{Cellular Users}
\newacronym{LUs}{LUs}{LEO Users}
\newacronym{GUs}{GUs}{GEO Users}
\newacronym{DRL}{DRL}{Deep Reinforcement Learning}
\newacronym{GPS}{GPS}{Global Positioning System }
\newacronym{HAPS}{HAPS}{High Altitude Platform Stations}
\newacronym{FDD}{FDD}{Frequency Division Duplexing}
\newacronym{FL}{FL}{Federated Learning}
\newacronym{QoE}{QoE}{Quality of Experience}
\newacronym{GBS}{GBS}{Ground Base Station}
\newacronym{OCC}{OCC}{Orthogonal Cover Codes}
\newacronym{TSGs}{TSGs}{Technical Specification Groups}
\newacronym{SI}{SI}{Study Items}
\newacronym{WI}{WI}{Work Items}
\newacronym{TR}{TR}{Technical Report}
\newacronym{eMTC}{eMTC}{enhanced Machine Type Communication}
\newacronym{HARQ}{HARQ}{Hybrid Automatic Repeat Request}

\usepackage{multirow}
\usepackage{array,boldline, makecell,booktabs}
\usepackage{longtable} 

    \usepackage[group-separator={,},group-minimum-digits=4]{siunitx}

\begin{document}
%
\title{A Tutorial on Non-Terrestrial Networks: Towards Global and Ubiquitous 6G Connectivity}
%
%
%

\author{Muhammad Ali Jamshed, Aryan Kaushik, Sanaullah Manzoor, \\Muhammad Zeeshan Shakir, Jaehyup Seong, Mesut Toka, Wonjae Shin, and Malte Schellmann
\thanks{
M. A. Jamshed is with the College of Science and Engineering, University	of Glasgow, UK (e-mail: muhammadali.jamshed@glasgow.ac.uk). \\ 
A. Kaushik is with the Department of Computing and Mathematics, Manchester Metropolitan University, UK (e-mail: a.kaushik@ieee.org). \\
Sanaullah Manzoor is with the School of Computing, Engineering and Built Environment, Glasgow Caledonian University, UK (e-mail: sanaullah.manzoor@gcu.ac.uk). \\
Muhammad Zeeshan Shakir is with the with the School of Computing, Engineering, and Physical Sciences, University of the West of Scotland, Paisley, UK (e-mail: muhammad.shakir@uws.ac.uk). \\
J. Seong and W. Shin are with the School of Electrical Engineering, Korea University, South Korea (e-mail: \{jaehyup, wjshin\}@korea.ac.kr). \\
M. Toka is with the Department of Electronics Engineering, Gebze Technical University, T\"{u}rkiye (e-mail: mtoka@gtu.edu.tr).\\
M. Schellmann is with Huawei Munich Research Center, Germany (email: malte.schellmann@huawei.com).
}
}

%
%

\markboth{}%
{Shell \MakeLowercase{\textit{et al.}}: Bare Demo of IEEEtran.cls for IEEE Journals}

\maketitle

\begin{abstract}

The International Mobile Telecommunications (IMT)-2030 framework recently adopted  by the International Telecommunication Union Radiocommunication Sector (ITU-R) envisions 6G networks to deliver intelligent, seamless connectivity that supports reliable, sustainable, and resilient communications. To achieve this vision, Non-Terrestrial Networks (NTN) represent a significant advancement by extending connectivity beyond the Earth's surface. These networks integrate advanced communication technologies that go beyond conventional terrestrial infrastructure, enabling comprehensive global connectivity across domains such as the Internet, Internet of Things (IoT), navigation, disaster recovery, remote access, Earth observation, and even scientific initiatives like interplanetary communication. Recent developments in the 3$^{\text{rd}}$ Generation Partnership Project (3GPP) Releases 17-19, particularly within the Radio Access Network (RAN)4 working group addressing satellite and cellular spectrum sharing and RAN2 enhancing New Radio (NR)/IoT for NTN, highlight the critical role NTN is set to play in the evolution of 6G standards. The integration of advanced signal processing, edge and cloud computing, and Deep Reinforcement Learning (DRL) for Low Earth Orbit (LEO) satellites and aerial platforms, such as Uncrewed Aerial Vehicles (UAV) and high-, medium-, and low-altitude platform stations, has revolutionized the convergence of space, aerial, and Terrestrial Networks (TN). Artificial Intelligence (AI)-powered deployments for NTN and NTN-IoT, combined with Next Generation Multiple Access (NGMA) technologies, have dramatically reshaped global connectivity. This tutorial paper provides a comprehensive exploration of emerging NTN-based 6G wireless networks, covering vision, alignment with 5G-Advanced and 6G standards, key principles, trends, challenges, real-world applications, and novel problem solving frameworks. It examines essential enabling technologies like AI for NTN (LEO satellites and aerial platforms), DRL, edge computing for NTN, AI for NTN trajectory optimization, Reconfigurable Intelligent Surfaces (RIS)-enhanced NTN, and robust Multiple-Input-Multiple-Output (MIMO) beamforming. Furthermore, it addresses interference management through NGMA, including Rate-Splitting Multiple Access (RSMA) for NTN, and the use of aerial platforms for access, relay, and fronthaul/backhaul connectivity. 
\end{abstract}


\IEEEpeerreviewmaketitle

\vspace{-3mm}

%


\section{Roadmap to 6G and Role of NTN: Why NTN is vital for the evolution of 6G networks?\textcolor{red}{}}

In a world where 2.9 billion people remain without internet access, addressing the digital divide has never been more critical. This disparity is particularly pronounced among certain demographics; in ten countries across Africa, Asia, and South America, women are 30-50\% less likely to use the Internet than men. Although cell technology is the most widely used communication system, it faces significant challenges, especially in rural areas, even in developed countries. \ac{NTN} offer a promising solution to these challenges, introducing new ways to connect the unconnected and enhance global communication \cite{jamshed2022challenges,kaushik2024toward,jamshed2020iot}.

\ac{NTN} refer to wireless communication systems operating above the Earth's surface, utilizing satellites in \ac{LEO}, \ac{MEO}, and \ac{GEO}, as well as \ac{HAPS} and \ac{UAV} \cite{hosseinian2021review,rinaldi2020non,zheng2024energy}. These elements are crucial to achieve uninterrupted coverage and extend connectivity to remote areas lacking traditional \ac{TN} access. Currently, devices are classified into those connected to \ac{TN} and those connected to satellites \cite{araniti2021toward,kaushik2022non}. This means that users who need satellite connections must use an additional device in conjunction with their smartphone. However, in an integrated system, all mobile devices will integrate both terrestrial and satellite access. As technology progresses, satellites are expected to function as \ac{BS}. \ac{NTN} plays a crucial role in expanding global connectivity, supporting various industries, and advancing technological capabilities. Therefore, the interconnection and inter-operation between \ac{NTN} and \ac{TN} are of significant importance \cite{geraci2022integrating,benzaghta2022uav}.
\begin{figure}
	\centering
	\includegraphics[width=\columnwidth]{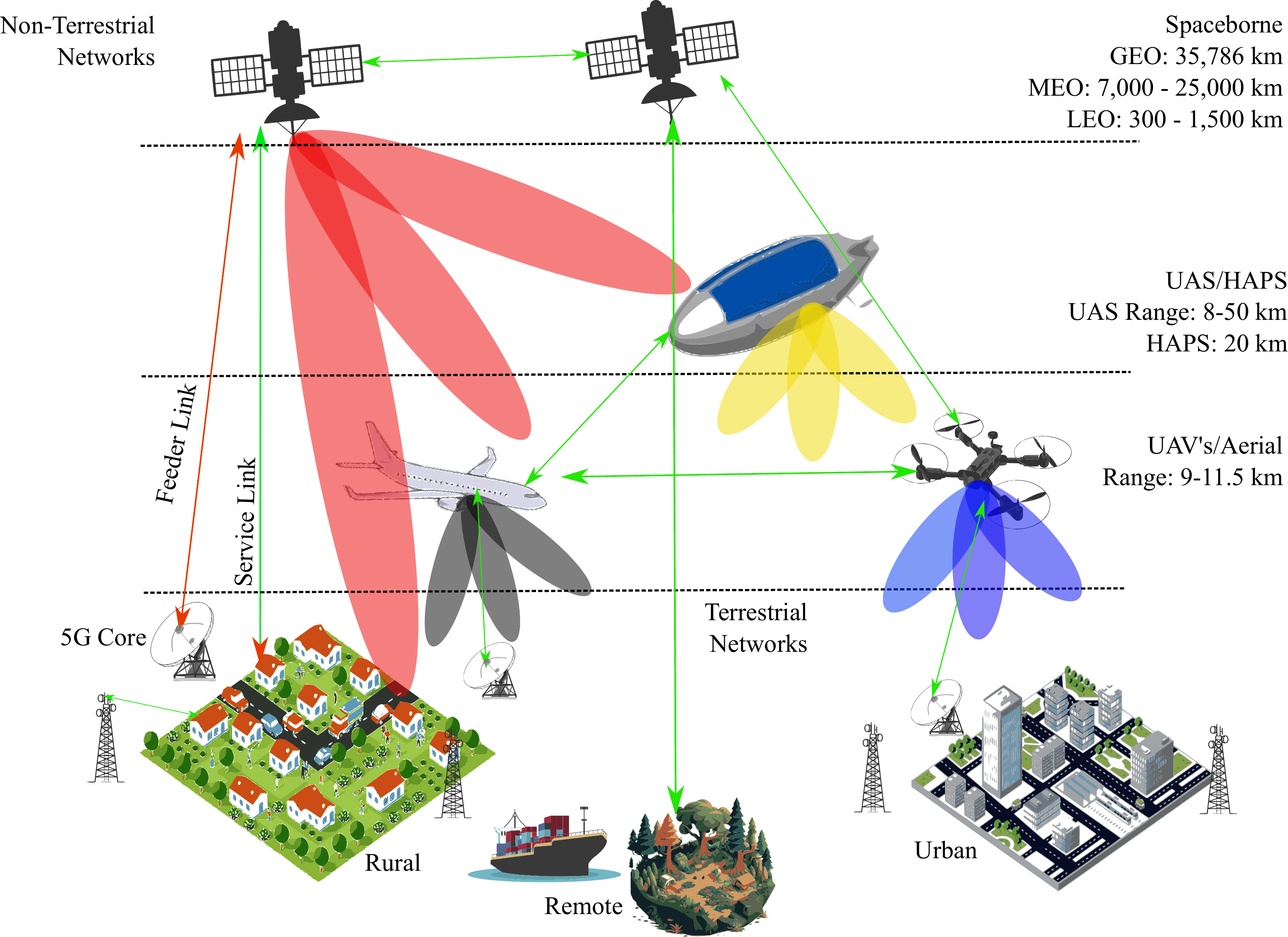}
	\caption{An illustration of convergence/co-existence of NTN and TN.}
	\label{fig:1}
\end{figure}

6G is expected to offer significantly superior connectivity compared to earlier generations, featuring higher data rates, reduced latency, and improved reliability. \ac{NTN} could supplement terrestrial 6G infrastructure by extending coverage to remote and under-served areas, where deploying traditional \ac{TN} is challenging or economically impractical \cite{singh2024towards,kaushik2024integrated,zhang2023satellite,katwe2024cmwave}. Within a 6G ecosystem, these \ac{NTN} will operate alongside \ac{TN} to provide seamless connectivity across various geographical regions, supporting initiatives such as the United Nations' 17 \ac{SDG}. The integration of \ac{NTN} within 6G networks will facilitate a wide range of new applications and use cases, many of which are extensions of current 5G applications, but have been limited by the performance constraints of existing networks. An illustration of the convergence and coexistence of \ac{NTN} and \ac{TN} is shown in Figure~\ref{fig:1}.

\begin{figure}
	\centering
	\includegraphics[width=\columnwidth]{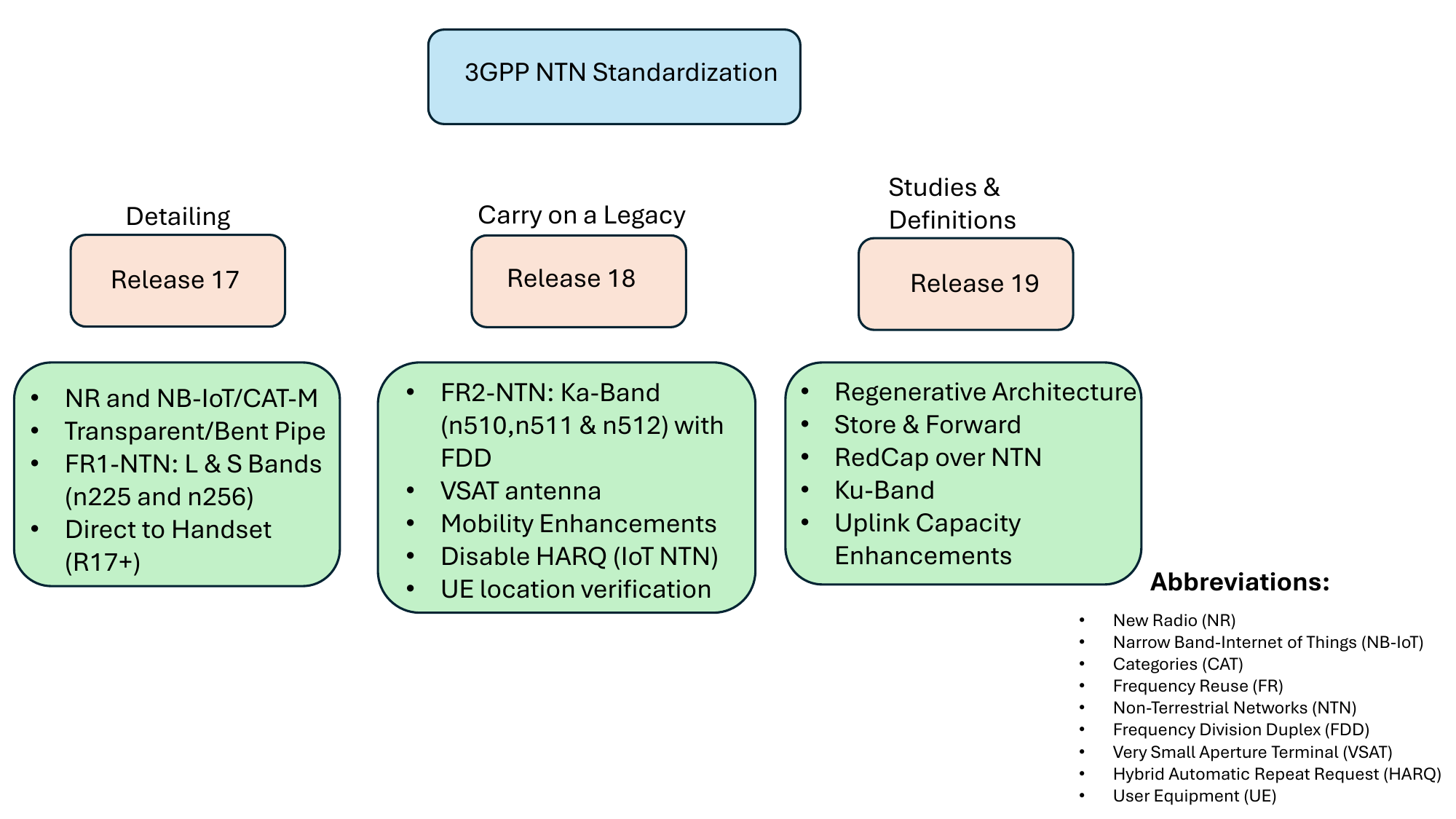}
	\caption{An illustration of 3GPP timeline indicating the start of NTN incorporation and future perspective.}
	\label{fig:21}
\end{figure}

\subsection{NTN Standardization in 3GPP}


The \ac{3GPP} is the global standardization body of cellular radio systems and their core networks. It was constituted in 1998 by seven telecommunication standard development organizations, aiming to develop the 3G mobile standards in an internationally aligned format to leverage creation of a standards eco-system that would facilitate global scale and create an enduring platform for its future evolution. Thanks to its success, it continued its standardization work to date, while keeping its original name – even though we are far beyond the 3rd Generation radio standard by now, peeking towards 6G. 

The work in \ac{3GPP} is structured into three \ac{TSGs}, which are defined as 
\begin{itemize}
    \item \ac{RAN}     
    \item \ac{SA} 
    \item \ac{CT} 
\end{itemize}

Each of the \ac{TSGs} is sub-structured into \ac{WGs}, where each WG focuses on a different level of the network or system, respectively, and they are simply numbered serially (i.e., RAN1 to RAN5 and SA1 to SA5). The work in the WGs is carried out in so-called \ac{SI} and \ac{WI}, where a \ac{SI} constitutes preparatory and pre-evaluation work for topics and aspects that are aimed to be covered by future releases of the standard. The outcome of a SI is summarized in a \ac{TR}, labeled with a specific 5-digit number in a 2-level format, e.g. 22.822. If a common consensus is reached to follow-up the work from the \ac{SI}, typically a \ac{WI} follows, where at the end a \ac{TS} is published carrying its own 5-digit number. 

The outcome of the standardization work is published in so-called Releases, where a Release provides a full functional description of the features and processes characterizing the cellular radio system and its core network. A Release consists of several TSs, which specify the normative requirements for the cellular radio system and are issued by the corresponding TSGs. A specific \ac{SI} or \ac{WI}, respectively, is usually defined for the duration of the working phase of a particular Release; hence, a \ac{SI} and a \ac{WI} referring to the same topic will typically be in subsequent Releases – though they may even appear in the same Release if the duration of the \ac{SI} in particular is defined significantly shorter than the duration of the Release’s working phase.

The complete roadmap of \ac{NTN} integration into the \ac{3GPP} cellular radio system from its beginning up to date is comprehensively described in \cite{3GPP_NTN_roadmap}, and will be summarized here in sufficient detail. First considerations on \ac{NTN} integration started in \ac{3GPP} Release 14 already, where the main driving factors were
\begin{itemize}
    \item Coverage extensions to areas with poor or without cellular coverage.    
    \item Services supported more efficiently by satellites, such as multicast/broadcast.
    \item Provision of a backup network in disaster zones with damaged cellular network.
    \item Higher resiliency of \ac{NTN}.
    \item Cost reduction through a unified radio interface for \ac{TN} and \ac{NTN}.
\end{itemize}

\subsubsection{\textbf{Release 15}}

Release 15 was the first to standardize the normative requirements for 5G \ac{NR}, which was initially focusing on \ac{TN} and various architectural options. The stage 1 \ac{TS} 22.261 included a requirement for 5G to support multiple access technologies, stipulating that the 5G system should be able to support mobility between supported access networks. However, due to time constraints, the standardization of satellite support was not included in Release 15. Despite this omission, TSG \ac{RAN} initiated a \ac{SI} entitled "Study on \ac{NR} to Support \ac{NTN}". This study focused on several key areas: 

\begin{itemize}
    \item \ac{NTN} use cases for \ac{eMBB} and \ac{mMTC} service. 
    \item Adapting the \ac{3GPP} channel model from Release 14 to accommodate \ac{NTN}.
    \item Providing a detailed description of deployment scenarios for \ac{NTN} while analyzing the necessary modifications to support satellite or \ac{HAPS} operations in \ac{NR}.  
\end{itemize}

The results of this \ac{SI} were summarized in TR 38.811. For \ac{eMBB}, \ac{NTN} use cases involve providing broadband connectivity to cells or relay nodes in under-served regions, in conjunction with terrestrial wireless or wireline access, though with limited user throughput. It also encompasses establishing broadband connections between the core network and cells in isolated areas, which is especially valuable for public safety applications. In addition, it facilitates the broadband connectivity between the core network and cells on moving platforms. For \ac{mMTC}, \ac{NTN} aims to ensure global connectivity between \ac{IoT} devices and the \ac{NTN} and to provide connectivity to a \ac{BS} that serves \ac{IoT} devices within a \ac{LAN}. 

\begin{figure}
	\centering
	\includegraphics[width=\columnwidth]{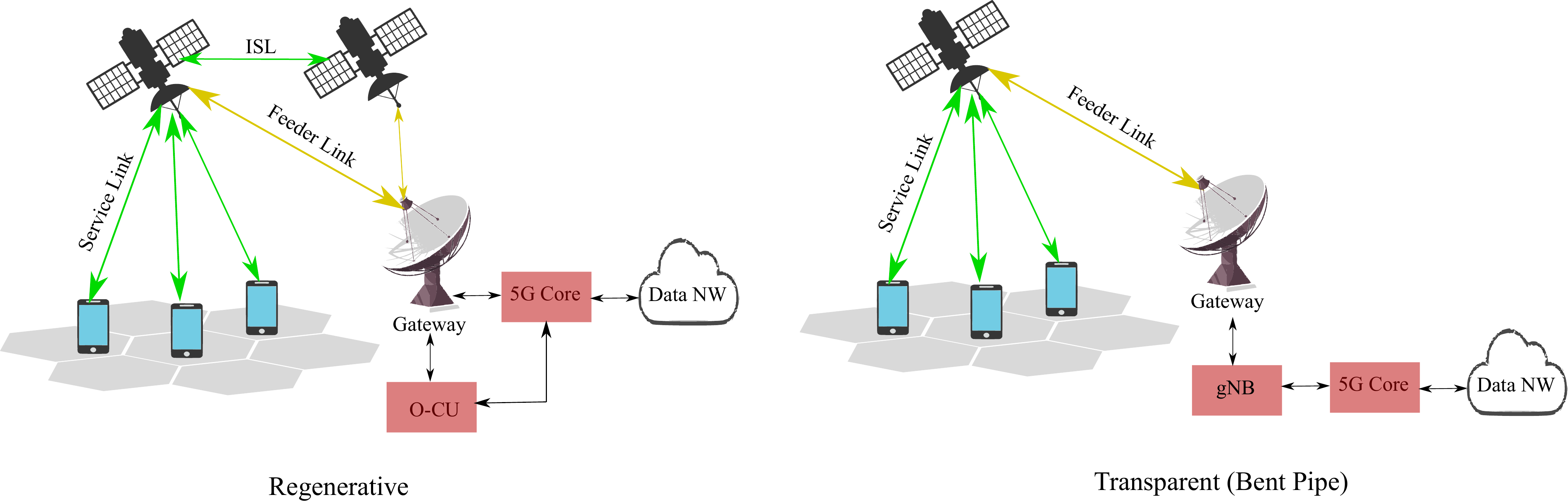}
	\caption{Transparent versus Regenerative payload.}
	\label{fig:2}
\end{figure}

The \ac{3GPP} adopted the common terms for characterizing the two main \ac{NTN} architectures: Regenerative and transparent payload, as illustrated in Figure~\ref{fig:2} and explained below.

\paragraph{Transparent payload}
A spaceborne or airborne platform that lacks on-board processing capabilities modifies the frequency carrier of the incoming uplink \ac{RF} signal, then filters and amplifies it before re-transmitting it on the downlink. In essence, this platform functions as an analog \ac{RF} repeater.

\paragraph{Regenerative payload}
A spaceborne or airborne platform that handles \ac{RF} filtering, frequency conversion, and amplification, along with on-board processing tasks, essentially operates as a \ac{BS}.

\subsubsection{\textbf{Release 16}}

In Release 16, TSG RAN and SA initiated one new SI each. The SI led by SA1, entitled "Study on using satellite access in 5G," analyzed 12 concrete use cases for using satellite access in \ac{NR}, which were summarized in TR 22.822. This SI assessed their conditions, impacts, interactions with existing services and features, and potential stage 1 requirements. These requirements included roaming between \ac{TN} and satellite networks, broadcast and multicast with a satellite overlay, \ac{IoT} via satellite networks, temporary use of satellite components, optimal routing or steering over satellites, satellite transborder service continuity, global satellite overlay, indirect connections through a 5G satellite access network, 5G fixed backhaul between \ac{NR} and the 5G core, 5G moving platform backhaul, 5G to premises, and satellite connections of remote service centers to offshore wind farms.

The RAN3 led SI focused on solutions for NR to support NTN, including support of NTN-TN service continuity and multi-connectivity scenarios for NTN-TN or for two NTN in parallel, which lead to TR 38.821. This SI built upon the key impacts identified in Release 15, examining the implications for \ac{RAN} protocols and architecture in greater depth and beginning to assess possible solutions. The investigation concentrated on satellite access via transparent \ac{GEO} and \ac{LEO} satellite networks, with \ac{HAPS}-based access being considered a special case of \ac{NTN} due to its lower Doppler and variation rates. The usage scenarios included pedestrians and users in vehicles, such as high-speed trains or airplanes. Reference scenarios evaluated included \ac{GEO} and \ac{LEO} satellites with both steering and moving beams, for both transparent and regenerative payloads. 
The RAN3 working group advised that the normative work should prioritize GEO-based satellite access with transparent payloads and LEO-based satellite access with either transparent or regenerative payloads. The outcome of both SIs yield proposals for the normative work to focus on with corresponding recommendations.

\subsubsection{\textbf{Release 17}}

Release 17 was the first release with normative requirements for \ac{NTN} in \ac{3GPP} specifications \cite{saad2024non}. In this release, the corresponding activities have also started to support \ac{NB-IoT} and \ac{eMTC} type of devices via NTN in 4G \ac{LTE}. 

The Release 17 WI "Stage 1 of 5GSAT" led by SA1 translated the findings from the corresponding Release 16 SI into stage 1 requirements, which were captured in TS 22.261. These requirements include the following: The 5G system must support service continuity between \ac{TN} and satellite networks owned by the same or different operators with agreements and must facilitate roaming. \ac{UE} with satellite access should support optimized network selection and re-selection to \ac{PLMN} with satellite access based on home operator policy, and must support mobility across various access networks. \ac{UE} must be able to provide or assist in providing their location to the 5G network, and the system must determine a \ac{UE} location to provide services according to regulatory requirements. There must be support for low power mobile \ac{IoT} communications and satellite links between the \ac{RAN} and core network, accommodating satellite backhaul latencies. Support for meshed connectivity between satellites with \ac{ISL} is required, as well as the selection of communication links based on \ac{QoS} fulfillment. 

The SA2 led WI focused on the integration of satellite components in the 5G architecture, resulting in the three specifications documents TS 23.501 (System architecture for the 5G system), TS 23.502 (procedures for the 5G system) and TS 23.501 (Policy and charging control framework for the 5G system). 

TSG RAN led WI "Solutions for NR to support \ac{NTN}", where the purpose was to adapt the basic features of 5G \ac{NR} to match the characteristics of the satellite channel. The focus was solely on the transparent architecture, and it was assumed that the \ac{UE} are \ac{GNSS} capable, allowing them to obtain precise position information. As operating bands for satellite communication in the frequency range FR1 (i.e., 410 MHz – 7125 MHz), the \ac{FDD} bands n255 (L-band), operating the uplink at 1626 - 1660 MHz and the downlink at 1525 - 1659 MHz, and n256 (S-band), operating the uplink at 1980 - 2010 MHz and the downlink at 2170 - 2200 MHz, have been introduced. For \ac{HAPS}, it was concluded that the FDD band n1 (uplink: 1920 - 1980 MHz, downlink: 2110 - 2170 MHz) can be applied, allowing \ac{NR} \ac{UE}s as defined by TS 38.101-1 to support \ac{HAPS} without any additional changes. 

New solutions and extensions of existing protocols proposed by the \ac{WI} to support \ac{NTN} operations cover several areas. Among those are enhancements in timing, synchronization, and \ac{HARQ} to cover long round trip delay; mobility management to ensure seamless handovers between NTN and terrestrial networks, utilizing satellite ephemeris and common parameters for the Timing Advance (TA); switchover procedures for the service link (i.e., between the \ac{UE} and the NTN node) and the feeder link (i.e.,between the ground station and the NTN node).

\subsubsection{\textbf{Release 18}}


In Release 18, the RAN-led SI entitled "Study on self-evaluation towards the submission of the International Mobile Telecommunications (IMT)-2020 submission of the \ac{3GPP} satellite radio interface technology" evaluated the \ac{NR} functionalities of Release 17 that facilitate 5G via satellite, as well as the corresponding \ac{LTE}-based solutions for \ac{NB-IoT}. In this SI, three different usage scenarios were considered for the analysis in a rural environment: 
\begin{itemize}
    \item eMBB-s (Enhanced Mobile Broadband - satellite), 
    \item   HRC-s (High Reliability Communications - satellite)
    \item mMTC-s (Massive Machine Type Communications – satellite).  
\end{itemize}

Furthermore, SI "Study on requirements and use cases for network verified \ac{UE} location for \ac{NTN} in \ac{NR}" pointed out the need for network-based methods to verify the reported \ac{UE} location within large \ac{NTN} cells, taking into account regulatory mandates for public alerts, emergency communications, and legal interception. Network-based verification with an accuracy of 5-10 km was mandated, prompting normative work in Release 18.

The operation bands for satellite communication have been extended by the new \ac{FDD} band n254, which operates the uplink at 1610 - 1626 MHz (L-band) and the downlink at 2484 - 2500 MHz (S-band), while the maximum channel bandwidth supported for NR NTN in FR1 has been extended to 30 MHz. Moreover, the WI "NR NTN enhancements" led by RAN2 aimed to enhance features from earlier releases, such as coverage enhancements as well as improved NR uplink coverage by enabling repetitions in the control channel and bundling reference signals for channel estimation. It also defined Rx/Tx time measurements for network verified UE location and introduced enhancements for NTN-TN and NTN-NTN mobility and service continuity, such as broadcasting geographical TN areas with frequency information, improved conditional handover triggers, and satellite switch with re-sync. Furthermore, NR-NTN operation in the above 10 GHz bands (targeting frequency range FR2) was considered, resulting in the introduction of \ac{FDD} bands n510, n511, n512 in the Ka band, operating uplink in the 17 - 20 GHz range and downlink in 27 - 30 GHz range, with Rx/Tx requirements for \ac{VSAT} UE types and \ac{RRM} for electronically/mechanically-steered beam UEs.

In TSG SA, several new SIs were started. Among those, the SI "Study on 5G core enhancement for satellite access phase 2" focused on handling mobility management and optimizing power savings under conditions of intermittent coverage, while the SI "Study on security aspects of satellite access" explored security and privacy issues related to mobility management and power conservation amid discontinuous coverage. The SI "Study on support of satellite backhauling in 5G system" concentrated on the use of satellite backhaul for mission critical scenarios with HRC-s enabled by satellite edge computing and local data switch – which requires a regenerative payload for its realization. The WIs driven by TSG SA focused on satellite backhauling in 5G system, where the results were captured in TS 22.261 and TS 23.501 - 503.

\subsubsection{\textbf{Release 19}}

In current Release 19, TSG RAN leads the WI "\ac{NTN} for \ac{IoT} phase 3", which aims to achieve several objectives for 2025. The link is improved by supporting additional satellite payload parameters for satellite constellations operating in FR1-NTN and FR2-NTN, improving the uplink capacity and throughput of FR1-NTN by using overlaid repetitions based on \ac{OCC}, signaling the intended service area of a broadcast service via \ac{NR} \ac{NTN}, and supporting for the first time in a WI regenerative payloads, featuring 5G system functions on the NTN node. Furthermore, the support of reduced capability (RedCap) devices (e.g., handheld or \ac{IoT}) for NR NTN operating in FR1-NTN will be specified. 

TSG SA is leading several SIs in Release 19. The SI "Study on Satellite Access - Phase 3" focuses on use cases and requirements to further improve the 5G system by satellite. Key subjects to be investigated are store and forward (S\&F) satellite operation for delay tolerant communication services (enabling services for discontinuous satellite coverage), direct UE-satellite-UE communication without using any feeder link to route the communication signal through a ground station (yielding significantly reduced communication delays), \ac{GNSS}-independent operation (to enable satellite access to UEs without \ac{GNSS} receiver/no access to \ac{GNSS} services), and positioning enhancements for satellite access (\ac{3GPP} based methods for satellite-only access). The normative part of this work will be done in a corresponding WI (starting at a later stage), and results will be captured in TS 22.261. 

Further SIs led by TSG SA are as follows: An SI on integration of satellite components in the 5G architecture focuses on regenerative payloads, S\&F satellite operation and UE-satellite-UE communication. An SI on management aspects of \ac{NTN} investigates management capabilities to support new network architectures or functions for satellite regenerative payloads, considering various satellite constellations. Another SI delves into security aspects of 5G satellite access, and finally, an SI focusing on application enablement will explore application layer solutions for satellite access. Last but not least, a workplan to support the Ku-band (downlink: 10.7 - 12.75 GHz, uplink: 12.75 - 13.25 GHz or 13.75 - 14.5 GHz) has been agreed.

\subsection{NTN Evolution}

\ac{NTN}, particularly through \ac{D2M} technology, leverage satellites to provide cellular connectivity directly to standard mobile devices. This approach offers several significant advantages. Firstly, it allows users to utilize their existing devices, ensuring accessibility without the need for additional specialized equipment. This can be transformative in disaster scenarios, where traditional infrastructure might be compromised. For instance, SMS has proven to offer the best performance-to-resource ratio in emergencies, and with \ac{NTN}, SMS and other essential services can reach even the most remote areas reliably. 

Beyond emergency communication, \ac{NTN} unlocks vast opportunities for the \ac{IoT} on a global scale. Remote sensors, powered by \ac{NTN} connectivity, can monitor a plethora of environmental and infrastructural parameters in real-time. For example, earthquake warning systems can benefit immensely from \ac{NTN}, where sensors deployed in seismically active but isolated regions can relay crucial data instantaneously, providing timely alerts that can save lives and mitigate damage. This capability is not limited to natural disaster monitoring; agricultural sectors in rural areas can also utilize \ac{IoT} devices to optimize farming practices, from soil moisture sensors to weather monitoring stations, thereby enhancing productivity and sustainability \cite{jamshed2017multicore,jamshed2021cooperative}.

Furthermore, \ac{NTN} benefits extend beyond technological advancements and emergency response. In regions with low internet penetration, \ac{NTN} can serve as a catalyst for social and economic empowerment -- particularly among women. By providing reliable internet access to remote and under-served areas, \ac{NTN} can facilitate education, healthcare, and entrepreneurial opportunities. Women, who are currently underrepresented among the internet users from a global perspective, can gain access to online resources, educational tools, and networks that were previously out of reach for them, thereby promoting gender equality and fostering inclusive development.

\section{Integrated, intelligent and ubiquitous NTN \textcolor{red}{}}

The rapid advancement of wireless communication technologies has significantly influenced how we connect, conduct business, and obtain information. Beginning with simple voice communications and progressing to high-speed data transmission and the \ac{IoT}, wireless communication has become integral to contemporary society \cite{jamshed2020antenna,jamshed2019survey}. However, as the need for faster data speeds, enhanced coverage, and more efficient use of the spectrum increases, the conventional communication infrastructure is faced with challenges. On one hand \ac{NTN} is considered a promising solution to overcome these limitations. However, the limitations associated with the use of \ac{NTN} limit its successful adoption to provide wireless coverage to terrestrial users. A comparison of \ac{TN} and \ac{NTN} is provided in Table~\ref{tab:1ctnntn}. A useful solution lies in the integration of \ac{NTN} and \ac{TN} \cite{nguyen2024emerging}. 

The inherent limitations of ground infrastructure, combined with economic factors, frequently impede the deployment of \ac{TN} in remote or hard-to-reach areas, such as rural regions, deserts, and oceans. As a result, \ac{UE} in these undeserved locations often cannot access terrestrial services. Integrating \ac{NTN} with current terrestrial infrastructure presents a practical and cost-effective strategy to achieve seamless and extensive wireless coverage. This integration offers the potential to improve network scalability and maintain continuous connectivity in these regions.

In such scenarios, \ac{NTN} play a crucial role in increasing the capacity, coverage, and speed of existing land-based networks. They address the shortcomings of ground-based infrastructure in meeting the required reliability and widespread availability standards for future wireless applications. Recognizing the benefits of \ac{NTN} and aiming to capitalize on economies of scale, the \ac{3GPP} is examining the integration of satellite access into the 5G framework \cite{jamshed2024synergizing,jamshed2022feasibility,jamshed2024guest}. Furthermore, as the 5G ecosystem evolves and refines its specifications—particularly for \ac{eMBB} and \ac{uRLLC}—it facilitates the seamless incorporation of aerial vehicles into \ac{TN}. The advanced capabilities of 5G technology meet the critical communication and control requirements of \ac{UAV}, supporting reliable connectivity for essential beyond visual \ac{LoS} operations \cite{awais2023energy,jamshed2023green,awais2022enhancing}, which are vital for numerous applications. Emerging applications such as autonomous vehicles, precision agriculture, and various yet-to-be-discovered use cases are driving extensive research efforts towards 6G to achieve global connectivity. The demand for worldwide coverage, along with upcoming plans for large \ac{LEO} satellite constellations, will further elevate the importance of \ac{NTN} in future networks. Fully integrated \ac{NTN} are poised to provide ubiquitous connectivity, delivering services anywhere, anytime, thus having a profound socioeconomic impact \cite{ryu2024rate}.

\begin{table}[!t]
    \centering
    \caption{Comparison of TN and NTN}
    \label{tab:1ctnntn}
    \begin{tabular}{|l|l|l|}
    \hline
        \textbf{Characteristic} & \textbf{TN} & \textbf{NTN} \\ \hline
        Delay/Latency & Lower & Higher \\ \hline
        Cost & Lower & Higher \\ \hline
        Trustworthiness & Robust & Vulnerable\\ \hline
        Data Rates & Higher & Lower \\ \hline
        Availability & Unreachable in Remote Areas & Reachable in Remote Areas\\ \hline
        Coverage & Regional & Global \\ \hline
        Mobility Management & Frequent Hand Overs & Seamless Mobility \\ \hline
    \end{tabular}
\end{table}

Wireless networks are evolving towards demand-aware autonomous reconfigurable networks. \ac{AI} has been identified as a key 6G technological enabler to adapt the network to a fast-changing environment \cite{nauman2021reinforcement,nauman2019intelligent,meng2024federated,jamshed2023reinforcement,khan2024advancing,paul2024quantum}. Similarly to human-brain functioning, data-assisted \ac{AI} models can learn from previous experience and execute the same action when faced with a similar perceptual environment, reducing complexity and computational resources with respect to a pure optimization-based decision-making \cite{haider2018performance,zhang2024collaborative,hu2023timely}.

Integrating \ac{NTN} into 6G systems introduces additional complexity to an already intricate infrastructure, marked by high-density \ac{BS} deployments and a rapidly growing number of terminals with diverse \ac{QoS} needs and unpredictable mobility patterns \cite{gao2022dynamic,gao2022virtual}. Traditional model-based approaches struggle to manage such complex networks due to their lack of mathematical simplicity and high computational demands. In this scenario, data-driven techniques offer a more efficient solution by learning and identifying patterns within complex algorithms, provided that they are trained on suitable datasets. Typically, offline training is improved through the continuous updating of \ac{AI} models with new data, which greatly improves the model's ability to adapt to evolving operational environments \cite{gao2020service,ryu2024traffic,gao2021energy}.

The \ac{3GPP} Release 18, marking the inception of the 5G-Advanced standard, includes considerations for a framework to incorporate \ac{AI} into the \ac{NR} air interface to improve network automation \cite{XingqinLin23}. Several study items have pinpointed key areas and use cases where \ac{AI} can significantly contribute. These include: 
\begin{itemize}
    \item Utilizing \ac{AI} for network management and orchestration.
    \item Enabling \ac{AI}-driven intelligence in the \ac{RAN}, with a particular emphasis on energy efficiency.
    \item Developing an \ac{AI}-native air interface that facilitates varying levels of collaboration between the gNB and \ac{UE}, covering aspects like \ac{CSI}, beam management, and positioning. 
\end{itemize}


In the following subsections, we will begin by exploring the applications of \ac{AI} in integrated 6G-\ac{NTN} networks and then proceed to discuss the associated challenges and limitations.

\subsection{\ac{AI} Innovations in Integrated 6G-\ac{NTN}}

The optimization and management of 6G-\ac{NTN} systems represent a key area with significant potential for \ac{AI}-driven solutions. Given the combination of ground-based stations and numerous satellites with overlapping coverage, it is crucial to closely coordinate both space and ground operations to ensure seamless service delivery to users. The network must be optimized to maximize resource efficiency while meeting service agreements with customers. \ac{AI} enables flexible and autonomous adaptation of the network to changes in wireless channels, fluctuations in traffic demand, and varying mobility patterns. Typically, \ac{DL} architectures are used to replicate complex algorithms, with the aim of striking a balance between performance and convergence speed. When testing in real environments is feasible, \ac{RL} can be used to experiment with various configurations through a trial and error process until optimal performance is achieved.

Proactively addressing network congestion and failures is crucial and requires continuous monitoring and analysis of network performance data. \ac{AI} tools designed for anomaly detection (such as identifying interference or link failures) and predicting network metrics (including demand, terminal trajectories, and congestion) have been shown to be beneficial for network management. These tools enable corrective actions to be taken before users' \ac{QoS} begins to degrade. This is particularly important in \ac{NTN} scenarios, where spectrum congestion is common due to the increasing number of \ac{LEO} satellites, making interference events more likely. Link congestion is another characteristic of \ac{NTN}, as a single satellite's wide coverage area of a single satellite includes significantly more users than a terrestrial \ac{BS}. Furthermore, repeated passage of satellites over the same areas can be leveraged, where satellite \ac{AI} local models are collected in a cloud through optically interconnected ground stations and updated based on shared data. \ac{ISL} can also be advantageous in developing cluster-based models with nearby satellites, using the experiences of neighboring satellites to enhance the general strength of the model.

\ac{NTN} offer consistent global coverage, which can be utilized for measurement and data collection purposes. However, the volume of data generated can be substantial, potentially overwhelming the on-board data processing and storage capacities, as well as saturating the downlink from the satellite to ground-based data centers. \ac{AI} tools have demonstrated significant potential to process these data for purposes such as feature extraction and change detection, thus easing the demands on on-board resources and minimizing the amount of data that needs to be transmitted to data centers. Furthermore, repeated passage of satellites over the same regions can be leveraged, with satellite \ac{AI} local models collected via optical interconnections at ground stations and updated using shared experiences. \ac{ISL} can also play a role in developing cluster-based models among nearby satellites, allowing the knowledge gained from neighboring satellites to be used to strengthen these models.

\subsection{Constraints of \ac{AI}-Powered \ac{TN}/\ac{NTN} Systems}

A fundamental feature of the \ac{O-RAN} architecture is its native design to support \ac{AI}-driven \ac{RAN}. \ac{RRM} can be optimized in near-real time (less than one second) to ensure efficient configuration of gNB radios, and in non-real time (more minutes or hours) to facilitate dynamic load balancing in \ac{TN} and \ac{NTN}, including the management of satellite beam \ac{EIRP} and onboard data routing. The \ac{O-RAN} \ac{ML} framework is structured to include: 
\begin{itemize}
    \item Multiple interfaces for data collection from various sources such as gNB components, \ac{UE}, core network, and other application functions.
    
    \item A host for \ac{ML} training to conduct both online and offline training sessions.
    
    \item \ac{ML} inference host that is primarily used for executing the \ac{ML} models, hosting the models, and providing output to the components involved in \ac{ML}-assisted solutions.
\end{itemize}

Firstly, the placement of these \ac{ML} model components greatly influences system performance, and \ac{NTN} presents specific challenges that are not yet fully understood. Various functional splits can be considered for \ac{NTN}, such as operating in a transparent mode with the entire gNB located on the ground, having only the \ac{RU} on board, placing both the \ac{RU} and \ac{DU} on board, or even having the complete gNB on board. Each scenario introduces distinct practical constraints for deploying and executing the \ac{ML} framework. Finding the optimal balance remains a challenge, considering factors such as long propagation delays (in contrast to \ac{TN}), potentially high numbers of \ac{UE} due to the broad coverage area of the satellite and, most importantly, the limited computational capabilities of satellite payloads. Furthermore, the dynamic topology of \ac{LEO} mega-constellations poses another significant hurdle for practical implementation.

Secondly, the effectiveness of \ac{AI} can be limited by the availability, volume, and quality of the data it relies on. Data are typically collected from various sources, such as \ac{UE} or network components, distributedly and on different spatial and temporal scales. As a result, modifications are required to the existing \ac{O-RAN} architecture, interfaces, and orchestration capabilities. The integration of new \ac{AI} generative models could help alleviate data scarcity by producing synthetic data sets that replicate real-world conditions. Furthermore, traffic modeling and forecasting play a key role in shaping \ac{ML} models for optimized resource management, which is critical to determining the most suitable approach for \ac{AI}-driven integrated \ac{TN} and \ac{NTN}.

Finally, many \ac{AI} algorithms discussed in the literature should mainly be viewed as theoretical (albeit useful) upper limits, because they often assume that a single entity operates the entire integrated \ac{TN} and \ac{NTN} system. Given the distinct technical architectures, historical evolution, regulatory frameworks, and geographical footprints of \ac{TN} and \ac{NTN}, these networks are likely to be organized differently from a business point of view. The primary methods for \ac{TN}/\ac{NTN} interconnection include roaming, multi-connectivity with a converged core, \ac{RAN} sharing, and their various adaptations. Each approach involves a different degree of network interworking, often limiting information sharing and leaving some data sources inaccessible due to confidentiality concerns. Studying the gap between these theoretical upper limits (assuming complete data knowledge) and more constrained \ac{ML} frameworks could offer valuable insights into the performance of \ac{AI} when applied to real-world \ac{TN}/\ac{NTN} architectural configurations and business models.

\section{AI and Deep Reinforcement Learning for Green 6G \ac{NTN} \textcolor{red}{}}
Optimizing NTN is crucial yet challenging, especially in dynamic, self-evolving settings, where it significantly impacts performance and efficiency \cite{nguyen2024emerging}. Optimization techniques helps \ac{MNOs} to select the best settings of the \ac{NTN} such as transmit level, beam-forming, and \ac{UAV} trajectory planning \& placement. This leads to overall network capacity \& coverage optimization, energy efficiency, resource allocation \& scheduling, and latency minimization. Various network optimization schemes have been introduced. Conventional \ac{NTN} optimization schemes are based on analytical simulations \cite{majamaa2022multi}, and mathematical models \cite{liu2021connectivity}. These schemes require prior context, knowledge about network parameters and environment to make decisions, e.g., network slicing based on users' requirement and network loads \cite{birabwa2021slice}, recovery networks during disaster situations and delivery task, etc. These aforementioned conventional \ac{NTN} optimization schemes have been considered for a long time, however, they have some intrinsic limitations.

The key limitations of conventional \ac{NTN} optimization schemes are as follows:
\begin{itemize}
    \item  Firstly, these schemes require prior context and knowledge about network environment \& settings which might be outdated or inaccurate due to the offline nature of algorithms, especially in mobile, dynamic, and highly uncertain conditions.
    \item Secondly, these approaches rely on predefined, fixed rules and models, which makes them non-adaptive to real-time network demands and changes.
    \item Thirdly, these schemes are designed by assuming static, and generalized network typologies and behavior; thus, they may be unable to tackle network heterogeneity and complexity.
    \item Furthermore, these schemes operate in reactive manner, which can lead to user's \ac{QoE} and \ac{MNOs} satisfaction minimization.
\end{itemize}

Therefore, \ac{AI}-based \ac{NTN} optimization schemes can address the limitations of conventional approaches, leading to more proactive, adaptive, and accurate network optimization  \cite{mao2021ai}.

\subsection{AI for \ac{NTN} Network Optimization} 

\begin{figure*}
	\centering
	\includegraphics[width=\columnwidth]{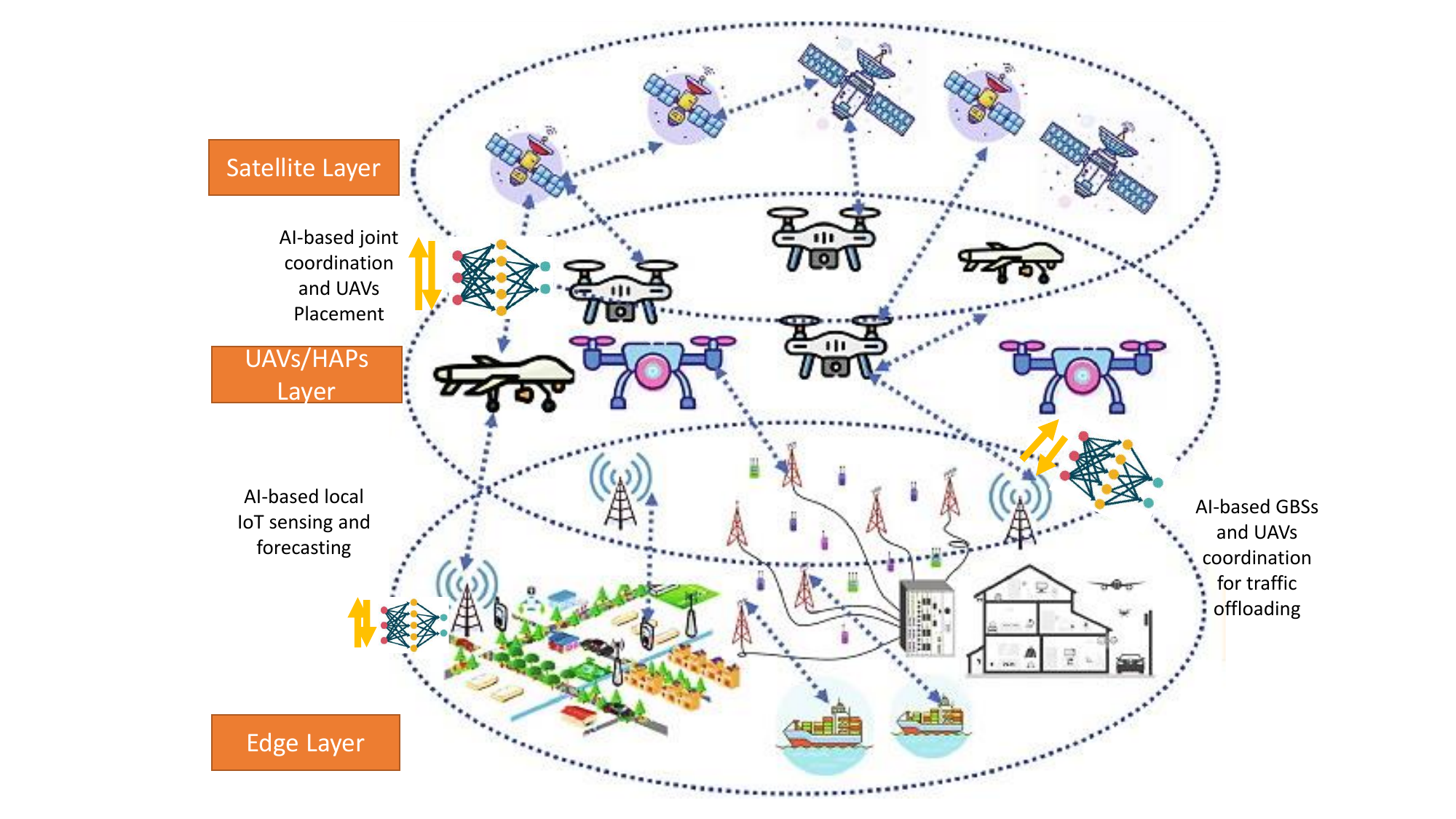}
	\caption{Overview of AI-based IoT NTN networks for edge-cloud computing based intelligent computation offloading}
	\label{fig_AI_IoT}
\end{figure*}

\ac{AI} can play a crucial role in enhancing the efficiency of \ac{NTN}. Given the limited resources available in NTN, AI techniques such as \ac{DL} and \ac{RL} are used to optimize network performance. These techniques predict traffic patterns \cite{shahid2024emerging}, manage power consumption \cite{qi2024key}, and improve resource allocation \cite{birabwa2022service}. In addition, AI is helping NTN dynamically adjust transmission power \cite{hosseinian2021review}, optimize routing, and schedule tasks more efficiently \cite{korikawa2024routing}, resulting in lower energy usage while maintaining reliable communication.  Fig. \ref{fig_AI_IoT} shows an overview of AI-enabled NTN. From Fig. \ref{fig_AI_IoT}, the architecture consists of three layers. The top layer uses LEO, MEO, and GEO satellites for backhaul links, connecting \ac{TN}, edge devices, and cloud data centers. The middle layer consists of \ac{UAV} and \ac{HAPS} located close to IoT devices, enabling local computation, real-time processing, and AI-driven analytics to minimize cloud dependency. The bottom layer includes \ac{GBS}, and \ac{IoT} devices, with AI managing resource orchestration and coordination with edge nodes for efficient computational offloading.

\subsection{AI-life cycle and Paradigms}
Basically, AI approaches learn from data and response to adapt dynamics of the environment. Recently, along with advances in computational hardware and the availability of open wireless and measurement data sets, applied AI algorithms have become more popular in \ac{NTN}. To date, several types of real trace-based \cite{boutiba2021radio} and synthetic \cite{rojas2024advanced} data sets are available for the \ac{NTN} research.  These data sets contain wireless environment statistics such as \ac{CSI}, user's association data, battery levels, \ac{GPS}, cell loads, link quality, and statistics are a few. 

Key AI technologies aiming to improve \ac{NTN} by performance using data-driven learning paradigms such as supervised, unsupervised, and reinforcement learning to tackle challenges in these networks. These paradigms use different types of data and training strategies to achieve different objectives. Here, we explain these paradigms in detail.

\begin{itemize}
    \item \textbf{Supervised Learning (Learning from Examples):} Under supervised learning, the algorithm comes with a data set containing input data that are linked to correct outputs, called labels. Hereby, this advice makes the model able to find out the patterns and connections between inputs and outputs as a result, becoming good at performing functions like classification and regression. Examples of supervised learning algorithms:
    \begin{enumerate}
    \item \textit{Linear Regression:} The linear regression is utilized to predict a continuous output relying on input features.
    \item \textit{Logistic Regression:} Logistic regression is used for binary classification tasks.
    \item \textit{Decision Trees:} Decision trees are a tree-like structure that is used for both regression and classification tasks.
     \end{enumerate}
\end{itemize}

\begin{itemize}
    \item \textbf{Unsupervised Learning (Unlabeled Data):} In unsupervised learning, you do not receive labels for the data. This implies that the system seeks patterns, clusters, or structures in the data. The model tries to uncover the hidden structure in unlabeled data by either grouping or transforming them.  Examples of unsupervised learning algorithms:
  \begin{enumerate}
    \item \textit{K-Means Clustering:} K-means clustering groups data points into a predefined number of clusters on the basis of feature similarity.
\item \textit{K-Nearest Neighbours (K-NN):} K-NN is a non-parametric method which is used for classification and regression based on feature similarities.  
\item \textit{Principal Component Analysis (PCA):} PCA is a technique used to simplify data generators by identifying key components.
   \end{enumerate}
\end{itemize}

\begin{itemize}
    \item \textbf{Reinforcement Learning (No Predefined Data):} In \ac{RL}, there is no predefined set of input-output pairs. Instead, the model teaches itself to make decisions by exposing itself to the environment and receiving feedback in terms of rewards or punishment. In reward-based learning, the system learns to take actions that eventually lead to maximizing cumulative reward. Some examples of algorithms are as shown below:
  \begin{enumerate}
    \item \textit{Q-learning:} This is a certain type of model-free RL algorithm in which agents learn how to make policies so as to maximize rewards attained.
\item \textit{SARSA algorithm:} This is an \ac{RL} algorithm that updates its action values based on the action taken and the final state reached.
\item \textit{Monte Carlo methods:} A set of algorithms that solve problems through random sampling and averaging the results to find a single number \end{enumerate}
\end{itemize}

\begin{figure*}
	\centering
	\includegraphics[clip,trim=0cm 0cm 5cm 0cm,scale=0.45]{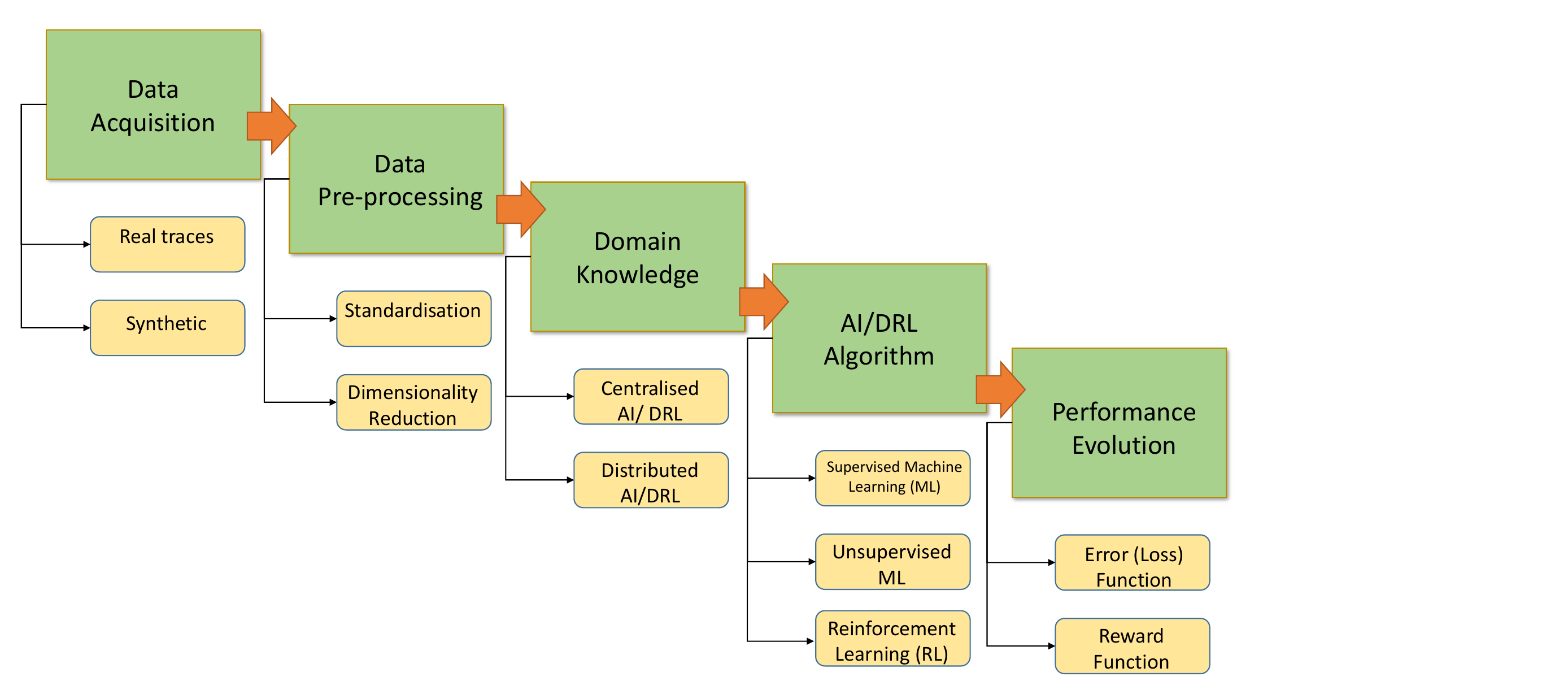}
	\caption{Overview of AI-life cycle for \ac{NTN} networks.}
	\label{fig_AI_Lifecycle}
\end{figure*}
Fig. \ref{fig_AI_Lifecycle} illustrates the \ac{AI},  \ac{ML} and \ac{RL} life cycle in a networked telecommunications \ac{NTN} context, structured in several key phases: 

\begin{itemize}
    \item \textit{Data Acquisition:} This involves collecting data from both real-world traces and synthetic sources.
\item \textit{(ii) Data Pre-processing:} Standardization and dimensionality reduction techniques are applied to clean and format the data.
\item \textit{(iii) Domain knowledge:} Based on the nature of the problem and the domain-specific knowledge is selected in a centralized or distributed AI paradigm. 
\item \textit{(iv) AI/DRL Algorithm Selection:} Centralized or distributed AI methods are used, including supervised, Unsupervised, and \ac{RL}. The selected algorithm is trained to optimize either a loss function (for supervised learning) or a reward function (for RL and DRL).
\item \textit{(v) Performance Evaluation:} The model’s performance is evaluated based on the precision of the predictions or the cumulative rewards.
\end{itemize}

\subsection{AI-based Tools \& Libraries for \ac{NTN} Research}

Here we will dive into some of the biggest platforms, libraries, tools created to support research for 6G \ac{NTN} communications.

\begin{itemize}
    \item \textbf{Network simulator 3 (ns-3):} AI support is an integral part of network simulator 3 (ns-3) \cite{yin2020ns3}. In addition to being a publicly available simulator of Internet systems, ns-3 offers customization possibilities through its modularity that allows integration of AI algorithms useful for 5G and 6G studies. ML-based network optimization tasks such as dynamic spectrum sharing, network slicing, or even self-organizing networks (SON) can be simulated using ns-3 supplemented with AI models. \textit{TensorFlow} and \textit{PyTorch} are key libraries that can be used in integrating these ML algorithms and thus can be run to imitate different \ac{NTN} case studies and scenarios.
    \item \textbf{AI Support in OMNeT++:} Libraries such as INET OMNeT++ \cite{cardenas2020tips} are capable of modeling more advanced forms of communication networks, including wireless and mobile networks. This would be possible through integration of AI algorithms so that they can automate changes within such networks in cases where system performance would not be adversely impacted especially when it comes to cellular/mobile communication usage patterns. The available support for artificial intelligence libraries like TensorFlow or Pytorch makes OMNeT++ suitable for simulating intelligent network behaviors (e.g. self-organizing network modeling).
\item \textbf{MATLAB support for AI and DL:} Using MATLAB and Simulink toolboxes to create ML models, algorithm development, and data analytics for various applications. MATLAB has DL Toolbox \cite{MFayekUnknownTitle2017}, RL Toolbox, and ML Toolbox that make their integration with 5G / 6G scenarios very convenient. There are several ways in which MATLAB can help: \textit{Beam-forming Optimization} – Using DL techniques can enhance beam-forming optimization massive MIMO systems: MIMO is explained as Multiple Input Multiple Output.
\textit{Predictive Maintenance} – ML models may imitate degradation of networking elements and thus avoid their damage. This leads to better network reliability.

\item \textbf{TensorFlow (Python-based library):} TensorFlow is an open-source AI library developed by \textit{Google} \cite{SilaparasettyUnknownTitle2020}. It is widely employed in telecommunications research because it has the capacity to support large datasets that model complex neural networks, and also train RL agents for resource management.

Communication engineers and researchers employ TensorFlow for various case studies such as: \textit{Model 5G/6G level networks:} TensorFlow contains within its structure networks with the ability to predict changes in traffic intensities or predict this based on various factors; moreover, these systems understand how to manage radio frequencies more effectively, thus reducing time delays between transmissions.

\item \textbf{Keras and PyTorch (Python):} PyTorch and Keras in TensorFlow are different DL toolkits that are well suited for telecommunication applications because they are easy to use, adaptable to various needs, and include the latest artificial intelligence techniques \cite{NuguesUnknownTitle2024}. An example of such a case is Adaptive Modulation and Coding (AMC), where AMC, designed to optimize data transmission efficiency under varying channel conditions, can be modeled and optimized by our AMC-based PyTorch in conjunction with Keras models that are able to distinguish better between noise patterns present at different signal levels at reception from desired data patterns.
\item \textbf{Scikit-Learn (Python):} Scikit-Learn is a widely known and popular library of traditional AI algorithms used in 5G and 6G research for classification, regression, clustering, and dimensionality reduction \cite{PaperUnknownTitle2019}. \ac{NTN} researchers can use these tools to implement standard ML algorithms due to modular and simplicity.

These are key usage of Scikit-Learn in different cases:  \textit{Traffic Classification:} scikit-learn has the ability to classify traffic into different service categories (eMBB, URLLC, mMTC) to improve QoS.
\textit{Predictive Modeling:} Scikit-Learn library is good for forecasting network traffic or predicting handover events on mobile networks.
\item \textbf{Distributed AI using Ray:} Ray is an open source library created for distributed AI and RL \cite{JaniUnknownTitle2024}. As for 6G \ac{NTN} studies, it enables one to deploy AI models across several machines while helping out with simulation involving thousands of nodes within the big network scenarios that need real-time decision-making processes, a fact that has made it one of the Python ML frameworks.

\item \textbf{Reinforcement learning using OpenAI Gym:} OpenAI Gym \cite{BeysolowIIUnknownTitle2019} provides developers and reviewers of RL algorithms a toolkit helpful in developing and comparing them, and therefore it is highly applicable in 5G/6G tasks such as dynamic spectrum allocation, power control, resource allocation among users. For example, simulating different 6G settings, such as RL within Gym, would be helpful when training agents to make optimal decisions given changing network conditions.

\item \textbf{R-Studio Libraries for Data Visualization, Plotting, and Modeling:} R-Studio has powerful specially designed packages necessary for real-time analytics, big data handling, and building strong ML models essential for realizing the full potential of 6G \ac{NTN} technologies \cite{AitkinUnknownTitle2022}. Efficient network monitoring, optimization, and innovation in 6G networks are made possible through the incorporation of sophisticated visualization tools alongside predictive modeling techniques within R-Studio. For example, R-Studio has \textit{cret}, and \textit{forecast} for data modeling. \textit{ggplot2}, and \textit{plotly} are commonly used data visualization libraries.
\end{itemize}

\begin{table}[t]
\centering
\caption{Summary of AI and DRL schemes for Green NTN}
\label{Table:AI and DRL schemes for Green NTN}
\begin{tabular}{|p{2cm}|p{11cm}|}
\hline
\textbf{Reference} & \textbf{Contribution} \\
\hline
\cite{zhang2023intelligent} & Proposed a novel DRL framework to optimize power allocation and beam coverage in LEO satellite networks, reducing energy consumption by up to 30\%. \\
\hline
\cite{li2022multi} & Developed a multi-agent DRL algorithm to jointly optimize spectrum and power allocation across satellite, UAV, and terrestrial networks, improving energy efficiency by 25\% while maintaining QoS. \\
\hline
\cite{yang2024federated} & Introduced a FL architecture for distributed training of DRL agents across multiple NTN segments, enabling privacy-preserving optimization of network operations and reducing overall energy consumption. \\
\hline
\cite{sharma2024role} & Proposed an AI-based predictive maintenance system using DRL to optimize satellite operations and extend their lifespan, potentially reducing space debris and improving long-term sustainability of \ac{NTN}. \\
\hline
\cite{fang2022noma} & Developed a DNN approach for dynamic spectrum access in hybrid satellite-terrestrial networks, improving spectrum efficiency and reducing energy consumption by adapting to varying traffic demands. \\
\hline
\cite{kim2024cooperative} & Proposed a DRL-based adaptive beamforming technique for LEO satellite and HAPS networks, significantly improving energy efficiency and coverage compared to traditional beamforming methods. \\
\hline
\cite{nguyen2024emerging} & Provided a comprehensive review of AI and DRL applications in green 6G \ac{NTN}, identifying key challenges and future research directions in energy-efficient network design and operation. \\
\hline
\cite{rodriguez2024training} & Introduced a quantum-inspired RL algorithm for resource allocation in 6G \ac{NTN}, demonstrating potential for significant improvements in computational efficiency and solution quality. \\
\hline
\end{tabular}
\end{table}

\subsection{AI-based Schemes for Green \ac{NTN} Networking}

This section reviews several key contributions in the green \ac{NTN} networking, highlighting the advancements and challenges in optimizing energy consumption and enhancing network performance through AI and DRL. Table \ref{Table:AI and DRL schemes for Green NTN} summarise key contributions of recent works for green \ac{NTN} . In \cite{zhang2023intelligent}, authors proposed a novel DRL framework for optimizing power allocation and beam coverage in \ac{LEO} satellite networks. Their approach achieved up to 30\% energy savings compared to conventional methods, showcasing the potential of DRL in reducing the energy footprint of satellite communications. \cite{li2022multi} extended this idea to a multi-agent environment, where DRL was used to jointly optimize spectrum and power allocation across satellite, \ac{UAV}, and terrestrial networks. Their results demonstrated a 25\% improvement in energy efficiency while maintaining the 
\ac{QoS}.

In \cite{yang2024federated}, authors introduced a FL architecture that allows distributed training of DRL agents across multiple \ac{NTN}  segments, thereby enabling privacy-preserving optimization of network operations. This approach not only improved energy efficiency, but also safeguarded user privacy, a critical concern in 6G networks. Similarly, \cite{sharma2024role} on extending the operational lifespan of satellite constellations by developing an AI-driven predictive maintenance system using DRL. Their work has implications for reducing space debris and improving the sustainability of \ac{NTN}. In \cite{fang2022noma}, authors addressed the challenge of dynamic spectrum access in hybrid satellite-terrestrial networks by developing a DNN approach. Their solution improved spectrum efficiency and reduced energy consumption by adapting to the varying demands of traffic. \cite{kim2024cooperative} explored the use of DRL for adaptive beam-forming in LEO satellite and HAPS networks, achieving significant gains in both energy efficiency and coverage. A comprehensive survey reviewed the applications of AI and DRL in green 6G \ac{NTN}, identifying key challenges and potential research directions in energy-efficient network design \cite{nguyen2024emerging}. Finally,  \cite{rodriguez2024training} introduced a quantum-inspired RL algorithm for resource allocation, which demonstrated promising improvements in computational efficiency and solution quality for 6G \ac{NTN} .

These studies collectively underscore the critical role of AI and DRL in the development of energy-efficient and sustainable 6G \ac{NTN}. The research highlights the diverse applications of these technologies, ranging from resource allocation and spectrum management to predictive maintenance and adaptive beam-forming, all contributing to the overarching goal of green and sustainable 6G networks.

\section{AI and ML for \ac{NTN} trajectory, placement and use cases in \ac{NTN} \textcolor{red}{}}

In satellite-assisted NTN, planning and placing of the UAV trajectory is very challenging due to various obstacles and distances. In such a situation, the UAV tries to remain in the coverage area of the ground network while moving from one area to another, while the \ac{UAV} maintain their horizontal planar movements, yet their vertical trajectories require optimization as well. For autonomous \ac{UAV} deployments in various situations such as delivery tasks, disaster scenarios, or on-demand networks, these \ac{UAV} encounter various issues such as path planning and obstacle avoidance \cite{nobar2021resource}. Many studies such as \cite{hayat2016survey}, show that the performance of the \ac{UAV} connectivity is based on the line-of-sight (LoS) signal link parameters while channel quality depends mainly on the SNR. Most importantly, differences in coverage range between localization and communication should be taken into account during trajectory optimization in joint problems. To optimize \ac{UAV} routes, several research efforts have been made such as trajectory planning with transmission power control \cite{hayat2016survey}, trajectory optimization based on data rate requirement \cite{li2018, manzoor2020leveraging}, energy-aware trajectory planning \cite{gao2023aoi}, etc. 

\begin{figure*}[t]
	\centering
	\includegraphics[clip,trim=0cm 0cm 0cm 0cm,scale=0.55]{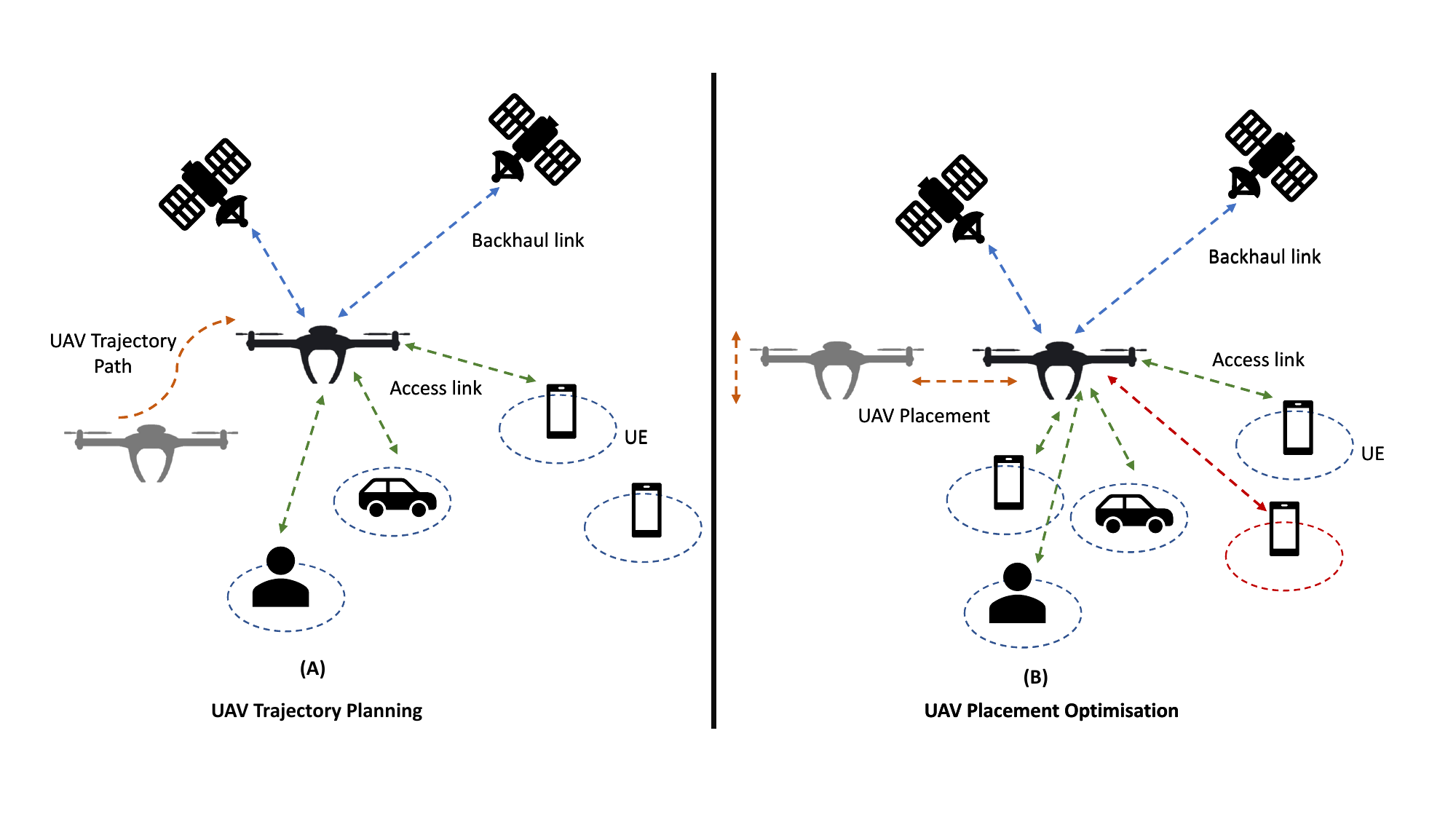}
	\caption{The process of \ac{UAV} trajectory planning and placement optimization.}
	\label{fig_UAVtrajectory-planning-nd-placement}
\end{figure*}

Fig. \ref{fig_UAVtrajectory-planning-nd-placement} shows two approaches to the flight path of the \ac{UAV} and the optimization of the position, underlining the importance of the \ac{UAV}’s path to assignment achievement and resource allocation. In \ref{fig_UAVtrajectory-planning-nd-placement} (A), the \ac{UAV} perform an initial deployment plan complete with paths and target locations, while adjustments are made on the right side in path and placement to improve performance and address dynamic environmental changes. The \ac{UAV} take irregular and longer paths (as depicted by dashed lines) like the orange path and green paths for the lower left \ac{UAV}, which consumes more energy and prolongs mission flights. In \ref{fig_UAVtrajectory-planning-nd-placement}(A), \ac{UAV} is optimizing horizontal and vertical positioning to maximize coverage and capacity.

\textbf{Limitations of Traditional \ac{UAV} Trajectory planning schemes:}

Path-planning schemes help \ac{UAV} navigate without interruption while traveling from a starting position to an endpoint (destination). Usually, \ac{UAV} might encounter various path interruptions, which are either environmental blockages or interruptions such as obstacles and objects. The path planning approaches of \ac{UAV} can be classified into various groups based on their objectives. The conventional \ac{UAV} trajectory planning scheme, such as \cite{chapnevis2023delay,rajendran2022machine, nemati2022non, farley2023generalized}, only considers blockage-aware path planning while ignoring overall flight-time minimization. In order to mitigate these issues, AI-based \ac{UAV} trajectory planning can be considered.

\begin{table*}[htbp]
    \centering
    \caption{Overview of machine learning-based \ac{UAV} trajectory planning approaches in \ac{NTN} networks}
    \label{tab:path_planning}
    \begin{tabular}{|c|p{4cm}|p{9cm}|}
        \hline
        \textbf{Ref. } & \textbf{Objective} & \textbf{Key Contribution} \\
        \hline
        \cite{wu2020optimal} & DRL-based multi-UAV path planning & Proposed a DRL-based multi-UAV path planning in a small squared grid region-based environment \\
        \hline
        \cite{bayerlein2021multi} & Known obstacles and grid-based cells & Proposed a new \ac{UAV} trajectory planning scheme with known obstacles and grid-based cells \\
        \hline
        \cite{hsu2020reinforcement} & DRL approach & Proposed a DRL-based obstacle avoidance and trajectory planning scheme \\
        \hline
        \cite{liu2022drl} & \ac{UAV} trajectory planning for IoT networks & Proposed a new \ac{UAV} trajectory planning and obstacle avoidance scheme in IoT networks \\
        \hline
        \cite{qian2022path} & Monte Carlo-based \ac{UAV} trajectory planning & Proposed Monte Carlo-based \ac{UAV} trajectory planning technique in wireless networks \\
        \hline
        \cite{zhang2022path} & Fixed wing-UAV based approach & Proposed a new \ac{UAV} path planning scheme for fixed wing-UAV in wireless environments \\
        \hline
        \cite{de2022q} & Q-learning-based multi-UAV path planning & Introduced a Q-learning-based multi-UAV path planning scheme for a 2D discrete grid map \\
        \hline
        \cite{dhuheir2022deep} & \ac{UAV} grid-based trajectory path & Suggested a \ac{UAV} path planning scheme using CNNs for grid-based trajectory path \\
        \hline
        \cite{zhang2022autonomous} & \ac{UAV} trajectory with random and dynamic environment & Proposed a new \ac{UAV} trajectory prediction in multi-obstacle scenarios with random and dynamic environment consideration \\
        \hline
        \cite{roghair2022vision} & Computer-vision-based \ac{UAV} path planning & Proposed a computer-vision-based \ac{UAV} path planning scheme for dynamic environment scenarios \\
        \hline
        \cite{zhang2023bionic} & Fusion-based \ac{UAV} & Proposed a new fusion-based deep neural network-based framework for \ac{UAV} path planning with obstacle avoidance and target tracking \\
        \hline
        \cite{xu2023real} & Camera input-based \ac{UAV} path planning & Proposed a camera input-based real-time obstacle tracking scheme for \ac{UAV} path planning and obstacle tracking in dynamic environment \\
        \hline
        \cite{seong2023multi} & Multi-UAV path planning for energy harvesting & Proposed a multi-UAV path planning approach for energy harvesting in known obstacles and cell grid environments \\
        \hline
        \cite{aslan2023goal} & CNN-based UAV path planning & Proposed a CNN-based \ac{UAV} path planning approach for static and dynamic obstacle avoidance \\
        \hline
    \end{tabular}
\end{table*}

\subsection{AI-based \ac{UAV} Trajectory planning schemes}
AI can improve the placement of \ac{UAV} and the joint optimization of network components such as antennas, RIS, relays, and routers. This optimization can lead to better coverage, increased capacity, better quality of service (QoS), and reduced costs and energy usage. Table \ref{tab:path_planning} shows an overview of recent contributions to \ac{UAV} trajectory planning for NTN networks. Table  \ref{tab:path_planning} provides an overview of ML-based \ac{UAV} trajectory planning approaches within \ac{NTN}. It highlights several techniques such as DRL, Q-learning, and Monte Carlo methods used to optimize \ac{UAV} path planning in various environments. 

Now, we discuss these works in detail. The paper \cite{wu2020optimal} introduces a DRL method for planning the paths of multiple \ac{UAV} within a small grid-based environment. This type of grid environment is often used to detach the \ac{UAV} flight region. In \cite{bayerlein2021multi}, a novel \ac{UAV} trajectory planning technique is presented that assumes known obstacles and uses grid-based cells to model the environment. The approach addresses how to navigate \ac{UAV} efficiently through such a predefined environment. In the work mentioned \cite{hsu2020reinforcement}, they use DRL to propose a combined obstacle avoidance and trajectory planning solution, focusing on ensuring that \ac{UAV} can avoid obstacles while planning efficient paths.

In \cite{liu2022drl}, authors present a new trajectory planning and obstacle avoidance technique specifically designed for IoT networks, where \ac{UAV} act as data collectors or relays. Similarly, the paper \cite{qian2022path} introduces a Monte Carlo approach to \ac{UAV} trajectory planning, focusing on applications within wireless networks, likely improving path optimization based on probabilistic methods. In \cite{zhang2022path} a trajectory planning scheme is tailored for fixed-wing \ac{UAV}. Unlike multi-rotor \ac{UAV}, fixed-wing \ac{UAV} have different maneuvering constraints, requiring specialized planning in wireless environments. In \cite{de2022q}, the authors utilize Q-learning for multi-UAV path planning in a 2D discrete grid map, focusing on the efficient movement of \ac{UAV} in a grid-based 2D space. In \cite{dhuheir2022deep}, authors applied CNNs to a grid-based trajectory planning approach, enhancing the path planning ability of \ac{UAV} in grid environments through the use of CNN-based techniques. The work in \cite{zhang2022autonomous} proposes a trajectory prediction technique that accounts for multiple obstacles in both random and dynamic environments, making planning more adaptable to unpredictable changes.

The authors in \cite{roghair2022vision} use computer vision for \ac{UAV} path planning, particularly in dynamic environment scenarios. The approach likely uses visual data from sensors to navigate in real-time. Similarly, the work in \cite{zhang2023bionic}, introduces a fusion-based deep neural network framework that integrates different data sources for \ac{UAV} path planning, emphasizing both obstacle avoidance and target tracking. \cite{xu2023real}, Camera input-based \ac{UAV} path planning scheme is proposed using camera input, aimed at improving \ac{UAV} path planning and obstacle tracking in a dynamic environment. In addition, the authors in \cite{seong2023multi} develop a multi-UAV path planning approach designed for energy harvesting, while operating in environments with known obstacles and cell networks. The work in \cite{aslan2023goal} proposes a CNN-based \ac{UAV} path planning method that deals with static and dynamic obstacle avoidance, improving \ac{UAV} navigation in complex environments.

\begin{table*}[t]
\centering
\caption{Overview of machine learning based \ac{UAV} placement approaches in \ac{NTN} networks}
\label{tab:summary_uav_placement}
\begin{tabular}{|p{1cm}|p{4.5cm}|p{9cm}|}
\hline
\textbf{Ref.} & \textbf{Objective} & \textbf{Key Contributions} \\
\hline
\cite{nguyen2019uav} & Bandwidth allocation and network capacity management & Proposed a novel \ac{UAV} placement scheme considering both bandwidth allocation and wireless backhaul capacity with QoS constraint. \\
\hline
\cite{dos2020energy} & Joint placement and transmission power optimization & introduced a Q-learning-based algorithm for joint \ac{UAV} placement and transmission power optimization to associate users while maximizing energy efficiency. \\
\hline
\cite{nouri2021three} & Edge-based computational task offloading & Proposed a \ac{UAV} placement to offload computational tasks from IoT devices based on the optimal number of \ac{UAV}. \\
\hline
\cite{shakhatreh2017efficient} & Connectivity for high-rise building users & Introduced a \ac{UAV} placement framework to provide connectivity to users in high-rise buildings. \\
\hline
\cite{shakoor2020joint} & User association and network capacity optimization & Proposed a 3D \ac{UAV} placement scheme by jointly optimizing user association, transmission power, and capacity. \\
\hline
\cite{alzenad20173} & Optimal number of UEs and \ac{UAV} prediction & Authors formulated an optimization problem to find the optimal placement of \ac{UAV} maximizing the number of UEs while minimizing transmission power. \\
\hline
\cite{wang2020joint} & Minimization of overall outage probability & Introduced a 3D \ac{UAV} placement scheme to minimize the overall probability of outage in the network. \\
\hline
\cite{alfaia2022resource} & Mitigating macro-cell overloading & Proposed a GRNN-based \ac{UAV} to mitigate macroc-ell overload in urban scenarios. \\
\hline
\cite{wang2020deep} & Predicting optimal position for user communication & Using GRU and CNNs to predict optimal position for \ac{UAV} deployment, improving overall user communication. \\ 
\hline
\cite{arani2020learning} & \ac{UAV} placement and interference minimization & A new mechanism is introduced to facilitate the three-dimensional deployment of \ac{UAV}, providing support to terrestrial networks during downlink congestion.\\ 
\hline
\cite{wang2024learning} & \ac{UAV} placement and interference minimization & Proposed a new approach to improve network throughput and meet real-time communication demands by deploying \ac{UAV}. The proposed method treats \ac{UAV} deployment as a computer vision problem and introduces a new method based on CNNs. \\ 
\hline

\end{tabular}
\end{table*}

\textbf{Limitations of Conventional \ac{UAV} Placement Schemes: }

Conventional \ac{UAV} placement techniques such as \cite{nguyen2019uav,pan20193d} operate in reactive manners and also rely heavily on centralized algorithms. These centralized algorithms assumed to have all information from the entire network. Furthermore, there are still scalability concerns, in addition to privacy issues and communication costs in such techniques \cite{manzoor2024mobility}.  In order to tackle these challenges, we can use AI-based learning approaches that can concern systematically and effectively providing better QoE.

\subsection{AI-based \ac{UAV} Placement Schemes}

Recently, advances in \ac{UAV} and satellites have introduced numerous opportunities for \ac{NTN} applications with flexible and low-cost deployment. In the existing infrastructure \text{5G/6G}, various UAV-based flying base station (FBS) deployments have been proposed in conjunction with ground base stations (GBS) \cite{hu2021uav,nguyen2019uav}. Table \ref{tab:summary_uav_placement} summarizes recent work on AI-based \ac{UAV} placement approaches in \ac{NTN}. In general, \ac{UAV} deployments can be classified as a centralized or decentralized (distributed) scheme based on their mode of control and operation \cite{nguyen2019uav,pan20193d}. The deployment of centralized \ac{UAV} poses various challenges in terms of scalability, privacy concerns, and communication overhead \cite{manzoor2024mobility}. Now, the section covers a detailed explanation of these contributions :

In \cite{nguyen2019uav}, authors focus on bandwidth allocation and network capacity management. The authors propose a novel \ac{UAV} placement scheme that takes into account both the bandwidth allocation and the wireless backhaul capacity while ensuring a QoS constraint. The key contribution is in designing a mechanism that balances the network's capacity and user demands by optimizing \ac{UAV} placement for bandwidth management, addressing the challenges of limited wireless backhaul in UAV-assisted networks. The paper \cite{dos2020energy} addresses the joint problem of \ac{UAV} placement and transmission power optimization. It introduces a Q-learning-based algorithm that optimizes the \ac{UAV}'s placement and transmission power, aiming to associate users while maximizing energy efficiency. This is an important contribution for energy-constrained \ac{UAV}, as it balances between coverage and power consumption using RL. This work in
\cite{nouri2021three} is focused on edge-based computational task offloading for IoT devices. The authors propose an optimal \ac{UAV} placement scheme to offload computational tasks from IoT devices to the \ac{UAV}, ensuring that the number of \ac{UAV} used is optimal. The contribution lies in efficiently distributing computational resources in edge computing, allowing IoT devices to save energy and processing power while maintaining performance.

In \cite{shakhatreh2017efficient} the authors address the challenge of providing connectivity for users in high-rise buildings. This paper proposes a \ac{UAV} placement framework designed specifically to enhance communication for users located in tall structures, where traditional terrestrial base stations may struggle to provide adequate coverage. The framework considers the three-dimensional nature of buildings and optimizes the positioning of the \ac{UAV} for seamless connectivity. The work \cite{shakoor2020joint} introduces a 3D \ac{UAV} placement scheme aimed at optimizing user association and network capacity. This approach jointly optimizes user association, transmission power, and network capacity, offering an integrated solution that enhances network performance by dynamically placing \ac{UAV} to respond to fluctuating user demands and network conditions. The paper \cite{alzenad20173} proposes a method to predict the optimal number of UEs and \ac{UAV} required in the network. The authors formulate an optimization problem where the objective is to place \ac{UAV} in a way that maximizes the number of UEs used while minimizing transmission power. This contribution is significant in balancing user demands and energy consumption in \ac{NTN} networks. The focus of \cite{wang2020joint} is minimizing the overall probability of outage in the network. The authors propose a 3D \ac{UAV} placement scheme that specifically aims to reduce the probability of service outages by optimally placing \ac{UAV} in locations that ensure continuous connectivity and reliable network service to users in challenging environments. Similarly, the paper \cite{alfaia2022resource} presents a ML-based solution, using generalized regression neural networks (GRNN), to mitigate macro-cell overloading in urban areas. The \ac{UAV} are placed strategically to offload traffic from congested macro-cells, especially in densely populated urban settings. The GRNN-based approach offers predictive capabilities for optimal placement of \ac{UAV} to relieve pressure on terrestrial networks.

In \cite{wang2020deep}, the authors use GRU and CNN to predict the optimal placement of \ac{UAV} to enhance user communication. This ML-based approach improves the network's ability to dynamically deploy \ac{UAV} in positions that enhance overall communication quality, offering a data-driven solution for real-time network optimization. The work in \cite{arani2020learning} proposes a mechanism for the placement of \ac{UAV} and the minimization of interference. The contribution is to provide support to terrestrial networks by deploying \ac{UAV} in a three-dimensional space, particularly during times of downlink congestion, thereby minimizing interference and enhancing network throughput. Distributed \ac{ML} such as \ac{FL} is a promising AI technology that enables multiple devices to collaboratively train ML models without sharing their sensitive data \cite{10147854, manzoor2024lightweight, 10233897, manzoor2022federated,manzoor2023fedbranched, mian2022value}. With regards to \ac{UAV} deployment, FL based placement schemes can be utilized to decide \ac{UAV} placement based on users' next-cell mobility information. The authors in \cite{wang2024learning} introduce a novel approach to improve network throughput and meet real-time communication demands by treating \ac{UAV} deployment as a computer vision problem. The authors propose a method using CNNs to optimize \ac{UAV} placement, which is a unique contribution that applies image processing techniques to solve network deployment challenges.

\section{Flying Platforms based Fronthaul/Backhaul in \ac{NTN} \textcolor{red}{}}

\begin{figure*}
	\centering
	\includegraphics[clip,trim=3.5cm 0cm 0cm 1cm,scale=0.68]{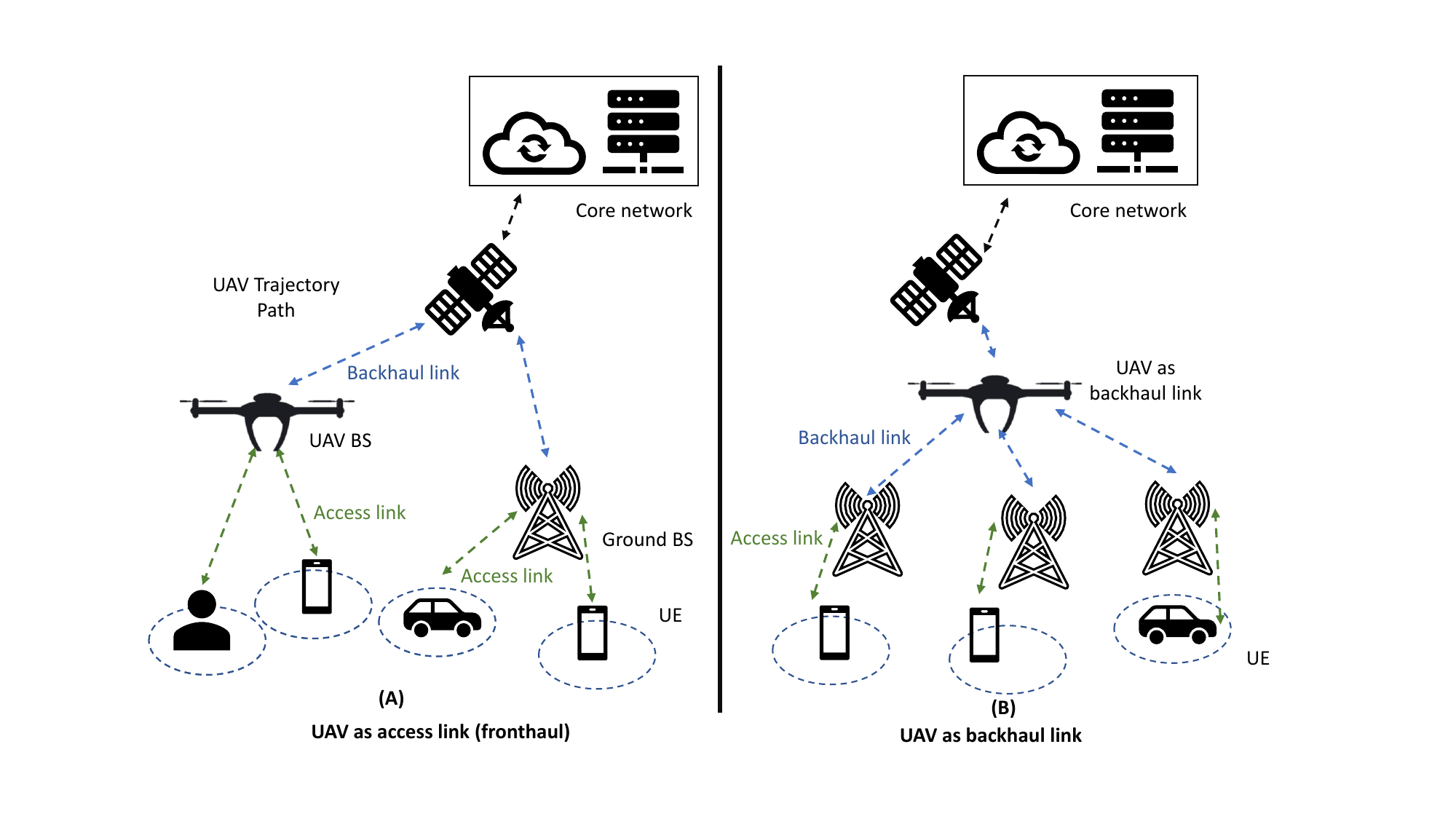}
	\caption{Overview of \ac{NTN} based fronthaul and backhaul links }
	\label{fig_AI_BK}
\end{figure*}

Integration of flaying platforms such as satellites, \ac{UAV}, and HAPS into 6G \ac{NTN} networks presents a variety of opportunities for backhaul and fronthaul access. This section discusses \ac{NTN} as fronthaul and backhaul network for the ground communication networks. Fig. \ref{fig_AI_BK} shows the front and backhaul communication architectures for \ac{NTN} networks.  In Fig. \ref{fig_AI_BK} (A) we have a \ac{UAV} as a fronthaul link that acts as an aerial base station through access links directly connecting with other users, as shown by green dashed lines. The \ac{UAV} connects to a central core network using a backhaul link (blue dashed line) showing a backhaul link. From Fig. \ref{fig_AI_BK}(A), the \ac{UAV} is acting as a relay network node within the central network infrastructure. Under this configurations, the \ac{UAV} becomes part of the fronthaul network that allows end users to communicate straight with a \ac{UAV} and communicate back to the network through a backhaul connection. In Fig. \ref{fig_AI_BK} (B), \ac{UAV} act as a backhaul link for the ground base stations. In this configuration, the ground BS connects users using access links (green dashed lines). In this setup, the \ac{UAV} helps UEs connect from different ground stations to the central core network. The dual role of \ac{UAV} as shown in Fig. \ref{fig_AI_BK} demonstrates the flexibility and significance of \ac{UAV}, especially in regions with scarce or limited connectivity such as remote locations, high-rise buildings or disaster situations where ground infrastructure is not available. 

\subsection{Impact of NTN Backhaul in 6G Communications}

Satellite or UAV-based backhauls connect \ac{PLMN} to the core network. In case of a satellite backhaul, \ac{RAN} can be shared between several networks, and thus this highlights its flexibility in deployment \cite{tezergil2022wireless}. On the other hand, this \ac{NTN} backhauling introduces some constant delay in few cases, as mentioned in the 3GPP release $18$ \cite{saad2024non}. This delay affects the overall communication infrastructure. In order to tackle this delay issue, \ac{MNOs} utilize satellite ephemeris information to counter the changes in network service. For example, a direct backhaul link can be used in case of \ac{GEO} satellite without inter-satellite links \cite{saad2024non, huang2024satellite}.

\begin{table}[t]
\centering
\caption{Recent contributions flying platforms for fronthaul/backhaul in 6G NTN}
\begin{tabular}{|p{3cm}|p{10cm}|}
\hline

\textbf{Reference} & \textbf{Contribution} \\
\hline
\cite{wang2024unmanned} & Provides a comprehensive overview of the integration of \ac{UAV} into 6G \ac{NTN}, focusing on front and back routes, with an emphasis on mobility management and energy efficiency. \\
\hline
\cite{mozaffari2019tutorial} & Explores drone-based base stations for emergency communications and proposes algorithms to optimize drone placement and link allocation. \\
\hline
\cite{abbasi2024haps} & Discusses HAPS and their advantages for fronthaul and backhaul in 6G \ac{NTN}, including wide coverage and stable positioning. \\
\hline
\cite{wu2019robust} & Presents fundamental work on optimizing UAV trajectories for communication, relevant to fronthaul and backhaul systems through energy-efficient path planning. \\
\hline
\cite{xu2020blockchain} & Investigates blockchain technology for secure spectrum allocation in UAV-assisted networks, applicable to fronthaul and backhaul resource management. \\
\hline
\cite{liu2020fast} & Proposes a ML approach for predictive deployment of \ac{UAV} based on traffic demands, with potential applications in fronthaul and backhaul efficiency. \\
\hline
\cite{hu2020energy} & Addresses energy limitations in UAV-based networks and proposes strategies to optimize communication and trajectories with multiple charging stations. \\
\hline
\cite{khan2022swarm} & Surveys fronthaul and backhaul technologies for 6G, including flying platforms and free-space optical communication for high-capacity links. \\
\hline
\end{tabular}

\label{table:literature_review_ronthaul/backhaul}
\end{table}

\subsection{Recent Research Directions Towards \ac{NTN} based Fronthaul and Backhual}
This section discusses the most recent trends \ac{NTN} based Fronthaul and Backhual. Table \ref{table:literature_review_ronthaul/backhaul} represents some recent work on flying platforms for fronthaul/backhaul in 6G NTN. In \cite{wang2024unmanned}, the authors provide a comprehensive review of how \ac{UAV} can be integrated into 6G \ac{NTN} to support both fronthaul and backhaul communication. The work in particular emphasizes mobility management and energy efficiency, addressing key challenges in maintaining reliable and sustainable communication links with aerial platforms. In \cite{mozaffari2019tutorial}, the authors explore the role of drone-based base stations in emergency communication scenarios. The study contributes by proposing algorithms to optimize drone placement and link allocation, ensuring that the communication infrastructure is quickly deployable and efficient in high-demand or disaster-stricken areas. The work in \cite{abbasi2024haps} focuses on HAPS and discusses their potential to provide front- and back-haul connectivity for 6G \ac{NTN}. HAPS offer advantages such as wide-area coverage and stable positioning, making them an attractive solution for delivering high-capacity, low-latency communication over large regions.

The work in \cite{wu2019robust} is fundamental for UAV trajectory optimization, presenting methods to improve communication efficiency by focusing on energy-efficient path planning. The study's contributions are relevant to both fronthaul and backhaul systems in 6G NTN, where UAVs need to maintain optimal communication links while minimizing energy consumption. In \cite{xu2020blockchain}, the authors investigate the application of blockchain technology to ensure the security of spectrum allocation in UAV-assisted networks. Using blockchain, the work addresses the secure management of resources in fronthaul and backhaul systems, ensuring that communication links remain efficient and secure from potential interference. The work in \cite{liu2020fast} proposes an ML-based approach for the predictive deployment of \ac{UAV}, with the model adjusting \ac{UAV} positions based on anticipated traffic demands. This dynamic method has implications for improving the efficiency of fronthaul and backhaul systems by aligning UAV deployments with real-time or predicted network requirements. In \cite{hu2020energy}, the authors address the challenge of energy limitations and propose new strategies to optimize both communication performance and \ac{UAV} trajectories using multiple charging stations. The proposed solutions are aimed at extending the operational time of the UAV and improving the reliability of communication networks in front- and back-haul scenarios. In \cite{khan2022swarm}, authors examine various technologies for 6G fronthaul and backhaul, including the use of flying platforms such as UAV and HAPS. The work also explores free-space optical communication, which can provide high-capacity links essential for 6G, offering a detailed overview of technologies and innovations required for future aerial communication networks.

In short, these works highlight the ongoing advancements in addressing the challenges of NTN-based fronthaul and backhaul networks.

\section{Next-Generation Multiple Access in \ac{NTN} \textcolor{red}{}}

\subsection{Multiple Access Techniques for Next Generation \ac{NTN}}

To achieve ubiquitous global and massive connectivity, \ac{NTN} is required to efficiently serve a large number of users in wide area coverage while addressing the increasing demand for high throughput. Given the limited radio resources available for \ac{NTN}, efficient resource utilization is crucial. To this end, advanced multiple access techniques are essential for achieving high spectral efficiency with global and massive connectivity.
In the following, we explore the candidates for \ac{NGMA} techniques suitable for \ac{NTN}, highlighting the potential to meet these stringent requirements.

\subsubsection{\textbf{Orthogonal Multiple Access}}

\ac{OMA}, such as \ac{FDMA} and \ac{TDMA}, assigns distinct frequency and time resource blocks to users or geographical areas.
By doing so, \ac{OMA} can manage signal interference and provide flexible satisfaction of specific \ac{QoS} requirements in \ac{NTN}.
For instance, in multibeam \ac{SATCOM}, a four-color beam pattern, dividing the usage bandwidth into two sub-bands and combining them with two orthogonal polarizations, is considered to manage inter-beam interference effectively. This ensures that inter-beam interference power remains \num{14} to \num{34} dB below the carrier signal \cite{perez2019signal}. 
In addition, beam hopping, which dynamically allocates beams to different geographical areas in a time-sequenced manner, is considered in multibeam \ac{SATCOM}. With beam hopping, interference between adjacent beams is reduced and the flexibility to allocate more radio resources to areas with higher traffic demand is provided in scenarios where \ac{QoS} varies across different regions and times. 
However, because \ac{OMA} divides and allocates usable resource blocks, simultaneously supporting a large number of users in \ac{NTN} is challenging.

\subsubsection{\textbf{Non-Orthogonal Multiple Access}}
Unlike \ac{OMA}, where a single carrier block is occupied by a distinct device at a given time, \ac{NOMA} allows multiple devices to use the same frequency block, thereby increasing spectral efficiency \cite{shin2017non,jamshed2021unsupervised,hu2021performance}. This is particularly beneficial for \ac{NTN}, where usable resources are more limited than in \ac{TN}. 
\ac{NOMA} can be classified into the \ac{PD-NOMA} and \ac{CD-NOMA}. In the \ac{PD-NOMA}, data is transmitted simultaneously at different power levels on the same frequency, while in the \ac{CD-NOMA}, data is transmitted concurrently by assigning different codes to users on the same frequency.
In particular, in the power domain \ac{NOMA}, the transmitter serves multiple users simultaneously by superimposing signals at different power levels. At the receiver end, interference signals from other users are decoded and removed via the \ac{SIC} technique before decoding the intended signal \cite{jamshed2022emission,jamshed2024electromagnetic,khan2022energy}. 
However, as the number of users within the network increases, the number of \ac{SIC} operations required at each receiver also increases proportionally, leading to higher receiver complexity. This complexity can limit \ac{NOMA}'s effectiveness in serving a very large number of users within the broadband coverage area of \ac{NTN}. Furthermore, if accurate \ac{CSIR} is not well guaranteed, residual interference signals remain after \ac{SIC}, resulting in error propagation that affects subsequent \ac{SIC} stages and limits the effectiveness of interference control.

\subsubsection{\textbf{Spatial Division Multiple Access}}
The spectral efficiency in the same frequency/time block can also be enhanced through \ac{SDMA}, which leverages the spatial \ac{DoF} provided by the \ac{MIMO} technique. In \ac{SDMA}, improved spectral efficiency is achieved with the beamforming technique that focuses the desired signal towards the corresponding user while reducing or nullifying interference signals by altering the phase or amplitude of the signals. The receiver treats interference signals as noise while decoding the desired stream, boosting spectral efficiency without adding receiver complexity. 
However, \ac{SDMA} relies on beamforming, making it more effective in user-underloaded systems where perfect \ac{CSIT} is available. \ac{NTN} often faces user-overloaded scenarios due to its broadband coverage capability. Also, obtaining perfect \ac{CSIT} is usually infeasible, particularly in satellite-based \ac{NTN}, due to factors such as long round-trip delays (around \num{250} ms for \ac{GEO} satellites and \num{30} ms for \ac{LEO} satellites) and the high mobility of satellites (around \num{7.5} km/s for \ac{LEO} satellites) \cite{10559954}.
Consequently, sufficient interference management in \ac{NTN} can be limited.

\subsubsection{\textbf{Rate-Splitting Multiple Access}}
In recent, \ac{RSMA} has emerged as a promising multiple access strategy to address these challenges \cite{10273395,mao2018rate, ryu2024rate}. A key idea of \ac{RSMA} is to split each user's message into common and private parts before performing beamforming at the transmitter. The common parts are combined into a single common message and encoded into a common stream with a shared codebook, making it decodable by all users. The private parts are individually encoded with dedicated codebooks, making them decodable only by the corresponding users.
On the receiver side, the common stream is first decoded, treating the private streams as noise, and subtracted from the received signal via SIC. Following this, each receiver decodes its private stream by treating the private streams of other users as noise. The users reconstruct their original messages with the information from the decoded common and private streams. 
\ac{RSMA}'s primary advantage lies in its flexible interference management, allowing it to partially decode interference while treating some of it as noise. By fine-tuning the balance between common and private parts, \ac{RSMA} offers high spectral efficiency, robustness to imperfect \ac{CSIT}, and scalability, while encompassing \ac{OMA}, \ac{NOMA}, and \ac{SDMA} as special cases. Thus, \ac{RSMA} is particularly useful in \ac{NTN} in that it can provide reliable service under user-overloaded and imperfect \ac{CSIT} conditions \cite{10273395, mao2018rate}. \ac{RSMA}, however, requires sophisticated signal processing to divide common and private parts, leading to higher computational demands at the transmitter \cite{shah2024multiple}. For practical and efficient real-world implementations, particularly given the limitations on onboard resources in \ac{NTN}, it is crucial to manage computational complexity in beamforming design carefully.

\begin{figure}
	\centering
	\includegraphics[width=\columnwidth]{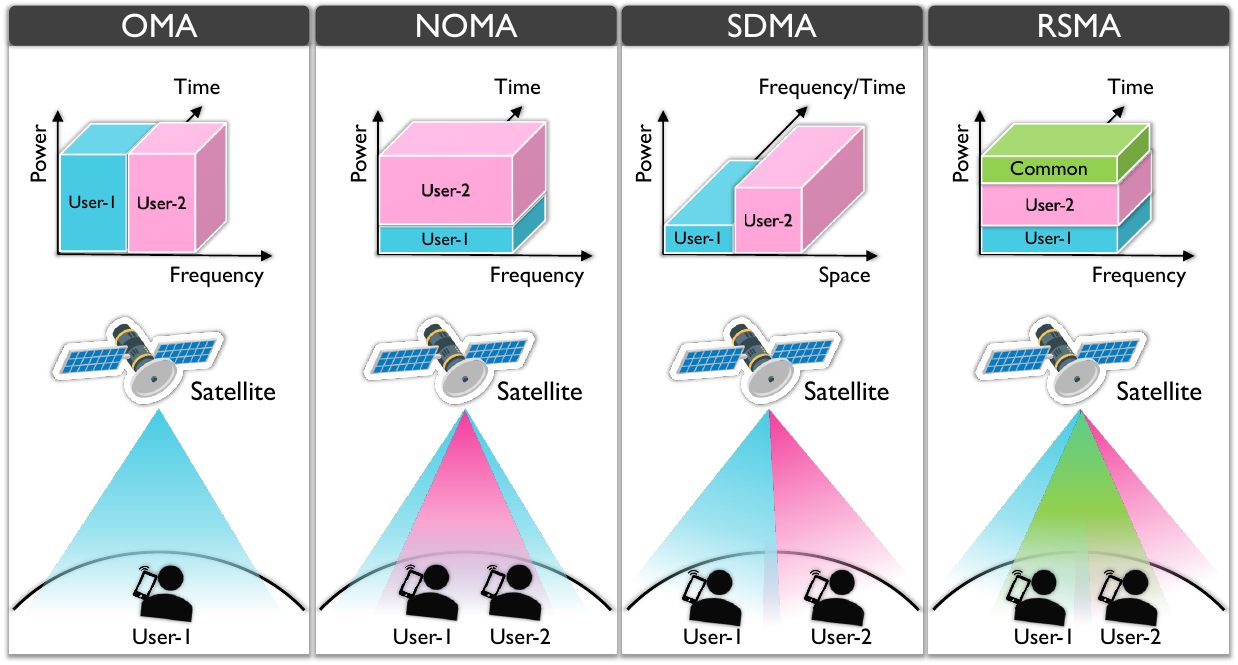}
	\caption{Comparison of resource distribution procedures for various NGMA strategies.}
	\label{Fig_1_im}
\end{figure}

Fig. {\ref{Fig_1_im}} summarizes the resource distribution procedures for various \ac{NGMA} strategies in a two-user case as a toy example, illustrating their distinct approaches to managing limited radio resources and co-channel interference in \ac{NTN}.

\subsection{Technical Challenges for Interference Management in \ac{NTN}: The Pivot Role of \ac{NGMA}}

\begin{figure}
	\centering
	\includegraphics[width=\columnwidth]{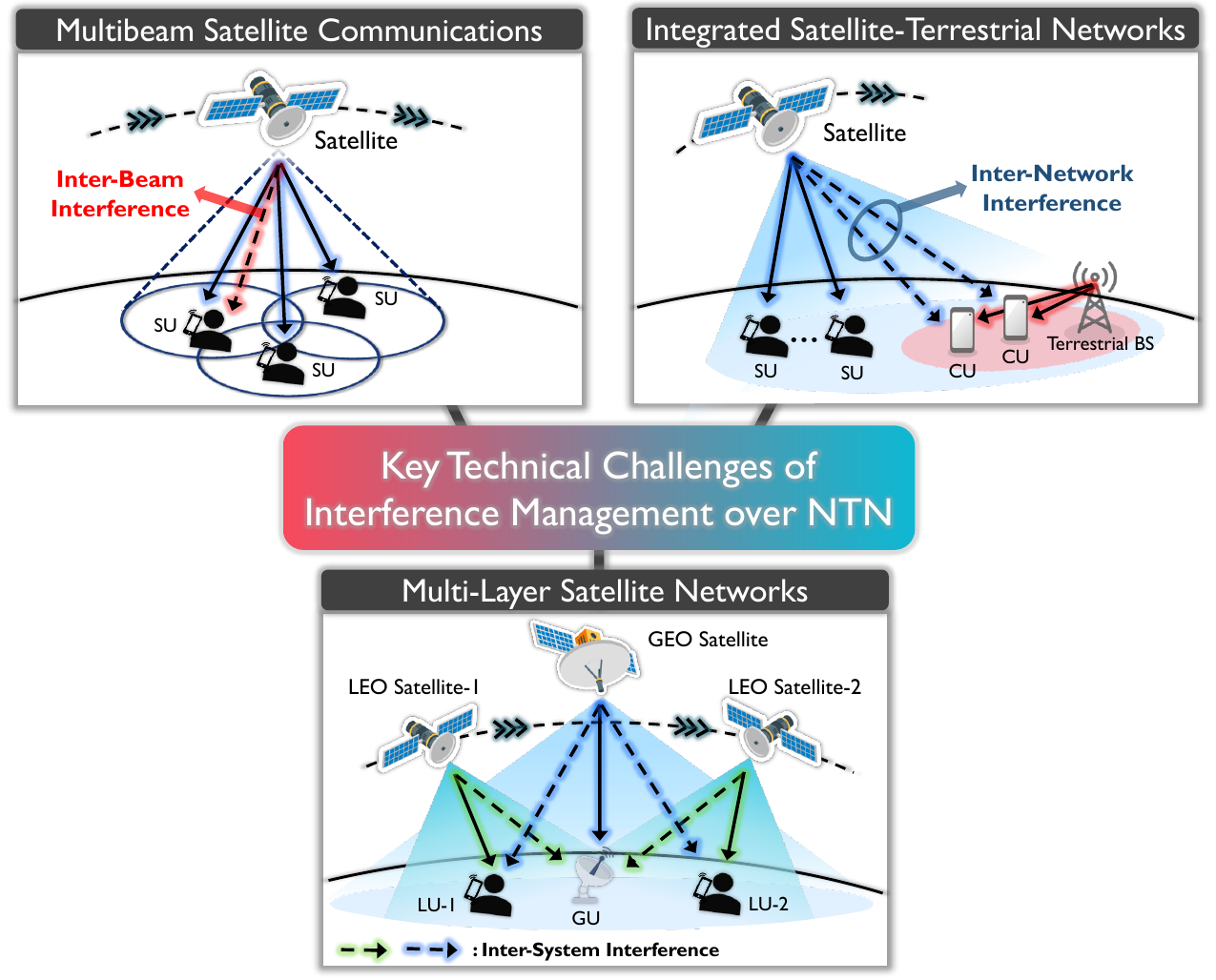}
	\caption{Technical challenges of interference management for various \ac{NTN} environments. SU: Satellite User, CU: Cellular User, LU: LEO User, and GU: GEO User.}
	\label{Fig_2_im}
\end{figure}

Managing interference in \ac{NTN} presents several technical challenges due to its unique characteristics. These include: i) intra- and inter-beam interference in multibeam \ac{SATCOM}, ii) intra- and inter-network interference in \ac{ISTN}, and iii) hierarchical interference in multi-layer satellite networks as illustrated in Fig. \ref{Fig_2_im}. 
\ac{NGMA} techniques have shown great promise in overcoming these challenges with flexible and robust interference management solutions. The following explores the technical challenges and the role of \ac{NGMA} in addressing these issues in various \ac{NTN} environments.

\subsubsection{\textbf{Multibeam \ac{SATCOM}}}

To achieve high throughput in multibeam \ac{SATCOM}, aggressive frequency reuse is indispensable, which leads to significant intra-beam and inter-beam interference. \ac{NGMA} techniques, such as \ac{RSMA}, have proven to be highly effective in managing these interference challenges. For instance, \ac{RSMA} has been used to address the max-min fairness problem in multibeam \ac{SATCOM}, ensuring uniformly good data rates in the extensive satellite coverage area \cite{yin2020rate_ms}. \ac{RSMA} manages interference issues under imperfect \ac{CSIT}, uneven user distribution per beam, and user overloaded scenarios.
Energy efficiency is another critical challenge in multibeam \ac{SATCOM} due to the power-hungry nature of satellites. To cope with this, \ac{RSMA} has been used to maximize the energy efficiency of multibeam \ac{SATCOM}  \cite{liu2023energy}. By improving spectral efficiency per unit power with \ac{RSMA}, effective power utilization is ensured even under channel phase perturbations caused by feedback delays.

The heterogeneous traffic demands within the satellite service areas due to the extensive coverage capability of \ac{SATCOM} pose another challenge.
The highly asymmetric traffic distribution can lead to mismatches between traffic demands and offered rates. To tackle this issue, \ac{RSMA}-based beamforming has been developed to minimize the disparities between traffic demands and offered rates \cite{cui2023energy, seong2023robust}. By flexible allocation of transmission power into common and private streams according to different traffic patterns among beams, \ac{RSMA} effectively satisfies traffic demands under both perfect and imperfect \ac{CSIT} conditions.

Although satellites are highly effective at delivering diverse multicast services, including video streaming, live broadcasting, and disaster alerts across extensive areas, the limited frequency/time resources make it challenging to meet the rapidly growing demands of next-generation wireless communications if distinct resource blocks are allocated for unicast and multicast services. To address this issue, \ac{RSMA}-based \ac{NOUM} transmission has been considered \cite{li2023non, Seong24}. By incorporating multicast messages into the common stream, which is decodable by all users within the network, \ac{RSMA} enables the simultaneous provision of unicast and multicast services without increasing the complexity of the \ac{SIC} operation at the receiver. Therefore, \ac{RSMA} helps overcome the limitations of scarce resources, improving both unicast and multicast communication services in multibeam \ac{SATCOM}. 

\subsubsection{\textbf{Integrated Satellite-Terrestrial Networks}}

The integration of satellite and terrestrial networks holds significant promise in providing extensive geographic coverage, especially in remote areas where terrestrial \ac{BS} infrastructure is unavailable or unreachable. In \ac{ISTN}, the management of inter-network interference and intra-network interference is paramount due to the shared resources between satellite and terrestrial \ac{BS}. As \ac{SUs} are typically located beyond the coverage area of \ac{TN}, interference from the terrestrial \ac{BS} to \ac{SUs} is generally not considered. However, interference from satellites to \ac{CUs} within the satellite coverage area is a significant concern that must be addressed.

To address these issues in \ac{ISTN}, coordinated and cooperative systems have been considered to manage intra-network and inter-network interference \cite{yin2022rate}. In the coordinated system, the satellite and terrestrial \ac{BS} share \ac{CSI} to manage inter-network interference. 
However, despite the shared \ac{CSI}, inter-network interference can still arise due to satellite-to-CU links.
The cooperative system, on the other hand, not only shares the \ac{CSI} but also facilitates data exchange between satellite and terrestrial BS via additional optical links, ensuring useful data transmission even through satellite-to-CU links that would otherwise cause signal interference. 
However, this introduces significant signaling overhead due to the need for data exchange between satellite and terrestrial \ac{BS}.

To mitigate data overhead in the cooperative system and inter-network interference in the coordinated system, an advanced coordinated scheme has been developed \cite{10266774}. This scheme, built upon the Han-Kobayashi strategy, a quasi-optimal approach for the two-user Gaussian interference channel, generates and transmits additional data from the satellite that contain messages intended only for \ac{SUs} but is decodable by both \ac{SUs} and \ac{CUs}. By doing so, \ac{CUs} can decode part of the interference from satellite-to-CU links, effectively reducing inter-network interference without the need for data sharing between the satellite and terrestrial \ac{BS}.

\subsubsection{\textbf{Multi-Layer Satellite Networks}}

To achieve the potential benefits of frequency reuse and multibeam techniques in \ac{SATCOM}, the coexistence of \ac{GEO} and \ac{LEO} satellites should be considered, as well as competition between large-scale \ac{LEO} constellations. However, this co-existence and spectrum sharing result in not only intra-satellite interference (i.e., inter-user interference from the intended satellite), but also additional hierarchical signal interference. That is, along with intra-satellite interference, \ac{GUs} encounter inter-system interference from \ac{LEO} satellites. At the same time, \ac{LUs} face both inter-\ac{LEO} interference from unintended \ac{LEO} satellites and inter-system interference from \ac{GEO} satellites. Advanced \ac{NGMA} techniques have shown promise in managing such interference issues.

Inter-system interference for coexistence networks of the \ac{GEO} and \ac{LEO} satellites can be managed with the cognitive radio method \cite{khan2023rate, ryu2024rate}. This method involves treating \ac{LEO} \ac{SATCOM} as a secondary network operating alongside the primary \ac{GEO} \ac{SATCOM}. To ensure the \ac{QoS} of the primary \ac{GEO} \ac{SATCOM} above a certain level, the interference level from the \ac{LEO} satellite to \ac{GUs} is restricted. Carrier assignment at the \ac{LEO} satellite can be performed to enhance the data rate of \ac{LUs} effectively.

The management of interference in the coexistence networks of a \ac{GEO} satellite and multiple \ac{LEO} satellites, distributed-\ac{RSMA}, which implements \ac{RSMA} in multilayer satellite networks, has been proposed \cite{xu2024distributed}. The centralized data processing gateway splits messages for \ac{GUs} into common and private parts, while messages for \ac{LUs} are split into super common, sub-common, and private parts. The \ac{GEO} common message and \ac{LEO} super common messages are combined and encoded into a super common stream, decodable by all \ac{GUs} and \ac{LUs}. Sub-common messages of each \ac{LEO} are combined and encoded into a sub-common stream with a codebook shared by corresponding \ac{LUs}. Private messages for \ac{GUs} and \ac{LUs} are individually encoded using codebooks known only by their dedicated users. Therefore, inter-system and inter-\ac{LEO} interference, as well as intra-\ac{LEO} interference, can be effectively managed thanks to the super common stream and sub-common streams, respectively, thereby enhancing overall network performance.

Using these key enabling \ac{NGMA} techniques, \ac{NTN} can address the complex interference challenges inherent in multibeam \ac{SATCOM}, \ac{ISTN}, and multi-layer satellite networks, thereby improving spectral efficiency, energy efficiency, and overall service quality, including service continuity. 
\section{RIS-Empowered 6G NTN: Potentials and Challenges \textcolor{red}{}}

\ac{RIS} has emerged as a groundbreaking technology for 5G and beyond wireless networks in the last decade. With distinctive features, such as the capability to complete duplex within a wide spectrum, low latency, flexible deployment, and the ability to deliberately manipulate radio propagation, the RIS offers promising solutions to meet demands in the ever-evolving landscape of wireless networks \cite{10559954,jamshed2024synergizing,ErtugrulGeorgeRIS}. The RIS is a planar meta-surface consisting of ultra-thin unit cells. Each cell comprises tunable components, such as PIN diodes, varactors, liquid crystals, memristors, etc., allowing us to manipulate instant electromagnetic waves. Those cells can be controlled using microprocessors to modify the phase and/or amplitude of the impinging wave. This brings in improved signal quality, extended service coverage, better interference management, and low energy consumption. Therefore, RIS technology has recently become a key enabler in the next generation of wireless communication. 

Initially, RISs have attracted the attention of academia and industry because of their nearly passive structures with lower power consumption compared to conventional ones such as relays. Passive type RISs make use of passive components that form an impedance-adjustable circuit to enable phase shift on the impinging signal \cite{7902179,9998527}. Those RISs do not use active RF components and introduce a negligible thermal noise. If a passive RIS consists of a high number of elements (unit cells), it is theoretically possible to achieve a high array gain. However, in practice, this is not possible due to the multiplicative fading effect when there is a strong LoS between the transmitter and the receiver. Additionally, the receiver could be located very far away from the RISs. Therefore, to overcome this issue, the active-type RIS concept has been proposed in the literature. This kind of RISs not only reflect but also amplify and forward the received signal. To do this, they use active reflection-type amplifiers, unlike passive-type RISs. Although they use active components, power consumption is still low compared to conventional ones such as relays \cite{9998527}.

Furthermore, a novel RIS concept named simultaneous transmission and reflection-RIS (STAR-RIS) has been proposed to further extend the coverage.  STAR-RIS enables 360$^{\circ}$ transmission coverage by reflecting and refracting the impinging signals \cite{9690478}. Unlike conventional RISs, STAR-RISs should have more metamaterial elements that allow induction, production, and radiation. Therefore, it is possible to flexibly determine beam steering through the indoor and outdoor receivers. Most recently, with advances in AI technology, the use of AI in the design of novel RIS structures has accelerated. For example, intelligent meta-surface stacking concepts (SIM), consisting of multiple cascaded meta-surface layers, have been investigated \cite{ErtugrulGeorgeRIS,StackedRIS1}. In a SIM, each layer can be digitally controlled in real-time with digital arrays, e.g., field-programmable gate array. The SIM can hierarchically control the energy of the electromagnetic wave that is reflected. In a layer, a unit cell (also known as meta-atom) acts as a re-programmable artificial neuron. The unit cell in the layer, where the electromagnetic wave is passing through, can be considered as a secondary source, and that cell can illuminate all the unit cells in the subsequent layer. All waves transmitted to the unit cell in the layer are superimposed; thus, all these waves act as the influx wave onto the unit cell in the subsequent layer. This process continues until the last layer. With the aid of a SIM, various signal processing tasks, for example image classification, can be handled in the electromagnetic wave domain \cite{ErtugrulGeorgeRIS}. Recently, the authors of \cite{StackedRIS2} have considered the application of the SIM concept in a holographic MIMO communication network. They showed that the SIM has the ability to reduce processing delay and overall energy consumption compared to MIMO transceiver designs.

In summary, owing to the remarkable benefits and flexible structures mentioned above, RIS technology opens many opportunities as a prominent phenomenon for 5G and beyond concepts. Therefore, many magazines and tutorial/survey-type papers have been published to attract the focus of researchers and strengthen the use of RISs in the forthcoming communication networks. Table \ref{RISMagSurveyPapers} shows an overview of the most prominent ones.

\begin{table*}[htbp]
    \centering
    \caption{Overview of existing most prominent magazine and tutorial/survey type RIS papers.}
    \label{RISMagSurveyPapers}
    \begin{tabular}{|c|p{1.5cm}|p{13cm}|}
        \hline
        \textbf{Ref. } & \textbf{Type} & \textbf{Key Contribution}  \\
        \hline
        \cite{MagRIS1} & Magazine  & Propose a programmable planar meta-surface, named Hypersurface tile, along with software-controlled architecture, and investigate its feasibility in the current networks infrastructures.   \\
        \hline
        \cite{MagRIS2} & Magazine  & Represent an overview for the RIS technology together with its prospective applications in wireless networks, and address the advantages and key challenges crucial for practical implementations.  \\
        \hline
        \cite{MagRIS3} & Magazine & Shed light on the practical use-cases of RIS technology by discussing the physical channel modeling, and introduce a new software-tool (graphical user interface (GUI)) for RIS-aided milimeter-wave systems valid for many wireless environments. \\
        \hline
        \cite{9690478} & Magazine & Provide an overview of a new RIS concept, namely STAR-RIS, and discuss its major differences between the conventional reflect-only RISs by elaborating hardware design and physical principles.  \\
        \hline
        \cite{ErtugrulGeorgeRIS} & Magazine & Delve into the unprecedented comprehensive world of RIS technology, especially elaborating hardware architectures, and unveil the interplay with emerging technologies, such as index modulation, integrated sensing and communication, and next generation multiple access.  \\
        \hline
        \cite{TutRIS1} & Tutorial /Survey & Introduce a detailed overview for RIS technology together with historical perspective, and explore its theoretical performance limits when applied in emerging wireless systems.  \\
        \hline
        \cite{TutRIS3} & Tutorial /Survey & Introduce a comprehensive overview for general smart radio environments empowered by RISs by addressing major concerns raised in the literature regarding underlying principles of hardware structures, working principle, consistency of the mathematical theories and electromagnetism, the most promising use cases and applications in wireless networks along with challenges, and current state of research. \\
        \hline
        \cite{TutRIS4} & Tutorial /Survey & Present a comprehensive survey for RISs regarding the applications and design aspects in forthcoming wireless networks, and focus on various performance metrics and approaches for the performance improvement. \\
        \hline
        \cite{TutRIS5} & Tutorial /Survey & Provide an overview of RIS technology to address the crucial challenges, such as reflection optimization, channel estimation, and deployment cases, and elaborate hardware architectures as well as various application scenarios.  \\
        \hline
    \end{tabular}
\end{table*}

\subsection{The Role of RISs in NTN}

One of the key objectives of RIS, for which it was originally created, is to improve availability, eliminate blind areas, and ensure that users are always connected on the ground. In the case of NTN-based communication, RIS has a lot of potential to enhance performance for cellular customers who depend on weak or absent connections to BS. In the following, we investigate separately the use of RISs in different aerial vehicle-assisted systems.   

\subsubsection{\textbf{UAV-assisted NTN}}

As demonstrated in \cite{jamshed2024synergizing}, UAVs that utilize RIS\textemdash a particular application of NTN\textemdash have effectively shown the advantages of RIS technology within NTN systems. On the one hand, RIS-mounted UAVs can complement conventional terrestrial infrastructures by expanding coverage and enhancing communication reliability, particularly for emergency incidents, temporary events, and/or remote areas, where establishing terrestrial infrastructures is challenging or expensive. To do this, the RIS mounted on the UAV reflects the received signal from terrestrial BS to the dedicated area \cite{AerialRIS}. However, RIS mounted on the surface of buildings can assist in UAV-based communication when the LOS environment between the BS and the UAV is harsh. In that case, the RIS generates an indirect path which boosts the received signal from the BS to the UAV. These applications allow for the continuity and quality of high-speed broadband service \cite{RISmeetUAV}. In light of this, the authors of \cite{RISUAVber} conducted a comprehensive performance analysis for a UAV-based communication network, where a RIS-mounted UAV maintains communication between the BS and a ground user. They provided new closed-form theoretical expressions for the outage probability, average bit error rate, and average sum-rate. In \cite{RISUAVoptim}, a URLLC system based on TDMA was considered, where a RIS-mounted UAV reflects the signal coming from an access point to IoT devices. Not only were the phase shifts of the RIS, but also the transmission duration of each IoT device and the location of the UAV were optimized by minimizing the total transmission power. In \cite{UAVRISdrl}, an energy efficiency maximization problem was considered in a RIS-assisted UAV network, where a RIS mounted in a building assists the UAV in serving a ground user. The 3D trajectory of the UAV and the phase shifts of the RIS were jointly optimized by implementing DRL algorithms, namely, double deep Q learning and deep deterministic policy gradient, to fully explore the mobility of the UAV and achieve high accuracy phase shifts.

Although there are many studies conducted in the literature, some of which are elaborated above, UAV-based communications with/without RISs may raise potential concerns about physical layer security (PLS). This is because illegitimate users (also known as eavesdroppers (Eves)) can capture signals transmitted by UAVs and RISs and can cause jamming effects. Therefore, PLS issues should also be considered when setting up RIS-aided and RIS-mounted UAV communications \cite{RISUAVpls}. One solution is that the RIS-mounted UAV can be located over the legitimate users with the help of the BS. On the other hand, phase shifts of the RIS are adjusted to boost the corresponding signals to the legitimate users. In addition, jamming attacks of Eves can be prevented by sending artificial noise signals which are directed by the RIS from the BS to Evas. To improve the PLS of a UAV-based network, in \cite{RISUAVplsRobust}, a terrestrial RIS mounted on a building was considered to strengthen the communication links of legitimate users and weaken Evas' links. To maximize average worst-case secrecy rate, a joint optimization problem was formulated considering the UAV's trajectory, phase shifts of the RIS, and transmit power of the legitimate users. Furthermore, the authors of \cite{RISUAVplsRobust2} considered the similar system of \cite{RISUAVplsRobust}, but also took into account multiple Evas to maximize the average secrecy rate.

\subsubsection{\textbf{HAPS-assisted NTN}}

HAPS are another cost-effective vehicle to serve remote areas, in which they have quasi-stationary mobility in the stratosfer relative to the Earth and operate around 20 km above. The coverage area of HAPS is about 50 km, and HAPS can also act as relay stations between satellites and UAV/BSs \cite{NTNRISalouini}. To establish a HAPS-based network, several criteria should be considered, such as low energy consumption, light payload, and reliable communication. At this point, the use of RISs has been regarded as one of the most promising solutions. For example, from a traffic backhauling point of view, remote area BSs maintain the traffic of ground users by establishing connections with a RIS-mounted HAPS. Thus, the RIS-mounted HAPS intelligently reflects the corresponding signals to a gateway station that has a connection to the core network. Additionally, reflecting the signals from one HAPS to another in a multi-hop relaying manner allows further enhancement of the coverage \cite{AerialRIS}. The authors of \cite{HAPSRis1} consider a network, where a RIS-mounted HAPS assists unsupported ground users to ensure the connection between the dedicated control station. To maximize the number of connected ground users and minimize the total power used by the control station and RIS, a new resource-efficient optimization problem was formulated. In \cite{HAPSRis2}, a novel network architecture consisting of aerodynamic HAPS mounted with RIS was introduced to ensure reliable communication between the BS and the ground users, particularly for emergency situations. A multi-objective optimization problem was proposed, designing phase shifts of the RIS, to maximize cascade channel gain. In this problem, mitigating the Doppler spread was also considered. Moreover, in \cite{HAPSRis3}, the authors provide a comprehensive performance analysis for two HAPS based communication networks; one consists of a RIS-mounted HAPS ensuring reliable communication between a satellite and a ground user, and the other addresses the use of a HAPS (acting as conventional decode-and-forward relay) and a building mounted on RIS in the same system setting. Closed-form mathematical expressions for prominent performance metrics, namely, outage probability, average symbol error rate of various quadrature amplitude modulations, and ergodic rate. The aerial RIS-assisted systems were shown to outperform compared to terrestrial RIS-assisted ones.

\subsubsection{\textbf{Satellite-assisted NTN}}

Given the recent technological progresses and requirements to achieve ubiquitous 6G connectivity, it is obvious that the exploitation of satellites is expected to be an indispensable pioneering approach. From the satellite type perspective, MEO satellites can offer wireless communication from 2G to 3G, while GEO satellites can only serve for 2G in terms of latency requirements. On the other hand, LEO satellites are able to provide 4G communication owing to low round-trip delay \cite{6GHyperCon}. Additionally, LEO satellites have many beneficial prospects, such as low-cost structure, offering high data rate, ensuring high reliability, providing high-precision positioning, and assisting high mobility applications. Therefore, with the recent advancements in satellite industry, it is envisioned that the use of LEO satellites is the most promising candidate for 5G and beyond communication networks \cite{LEOVLEO}. However, satellites face many challenges; for instance, GEO satellites suffer from high propagation delay increasing path-loss effect. MEO satellites depend on the trade-off between the coverage and delay. LEO satellites face severe Doppler effect, limited power, and short visibility. To combat many challenges encountered in satellite-assisted NTN and reinforce the integration of TN and NTN, the use of RISs has attracted great attention \cite{NTNRISalouini,jamshed2024synergizing,10559954}. For instance, the authors of \cite{Tekbiyik} explore improving the coordination between NTN and deep-space networks with the aid of RIS deployment. The performance behavior of RIS-assisted inter-satellite links is investigated in the presence of harsh environments, such as solar scintillation and misalignment fading. It is demonstrated that transmit power should be adopted according to satellite's position relative to the Sun, and novel beam alignment approaches should be developed to avoid environmental effects. In \cite{Tekbiyik1}, the authors consider terahertz band communication by quantifying misalignment fading and propose mounting RISs on LEO satellites to combat high path loss in terahertz-based inter-satellite network. In \cite{RIShstn}, RIS-aided hybrid satellite-terrestrial relay networks, where a RIS cooperates a BS to boost the satellited signals corresponding to the blocked users. An alternating optimization approach exploiting singular value decomposition together with uplink-downlink duality is proposed to minimize the total transmit power of the satellite and BS. Phase shifts of the RIS are also optimized using Taylor expansion and penalty function methods. On the other hand, the authors of \cite{RISsatUAVmultibeam} consider using multiple RIS-mounted UAVs to complement the communication between a satellite and ground user in a multibeam communication networks. To ensure the connectivity, each RIS-mounted \ac{UAV} hovers on each beam area. The authors investigated the impact of frequency reuse factor on the sum rate, because RISs may also amplify inter-beam interference as well as the desired signals. To demonstrate the performance behavior of the network, the closed form ergodic sum rate is derived in the presence of hardware impairments. In \cite{RISsatUAV}, a terrestrial RIS is considered to improve the signal strength of the ground user. In the network, a satellite transmits the ground user's signal to full-duplex UAV because of the blockages, then the RIS assists UAV for the re-transmission of corresponding signals. The closed-form outage probability is obtained for the end-to-end transmission in the presence of imperfect hardware and co-channel interference. The authors of \cite{RISmultiLEO} propose making use of a terrestrial RIS to mitigate the severe signal attenuation caused by satellite and terrestrial environment in a multi-LEO satellite network. It is considered that multiple LEOs can transmit the desired signals to a ground receiver within the visible time. An alternating optimization algorithm is applied to maximize the received signal power subjected to on-board power limits of satellites and modulus constraints of the RIS's phase shifts.

As seen in the studies above, there is a great deal of effort to improve signal quality between satellites and ground users with the aid of terrestrial and/or aerial RISs. However, efficient resource utilization is of vital importance in achieving high spectral efficiency. Therefore, investigating the integration of RISs in NGMA-based satellite networks has also received considerable interest in the literature. For example, the authors of \cite{STARrisNOMALEO} propose a novel signal constellation scaling and rotation in a terrestrial STAR-RIS-aided NOMA-based LEO satellite network. To achieve optimal performance of bit error rate, a joint maximum likelihood detector is considered; thus, a significant performance improvement is demonstrated via extensive simulations. In \cite{EErisNOMALEO}, an energy efficiency maximization problem is analyzed for a terrestrial satellite network based on RIS-assisted NOMA. In the network, a multi-antenna BS communicates with multiple NOMA users, and a LEO satellite transmits common information to multiple users. The proposed optimization algorithm for RIS-assisted NOMA is shown to outperform its counterparts. Likewise, the authors of \cite{EErisNOMALEO2} also consider the energy efficiency problem in a NOMA-based RIS-assisted LEO network in the absence of terrestrial BS communication. Furthermore, in \cite{CanIbrahim}, the use of terrestrial RIS in an RSMA-based uplink satellite network is proposed because RSMA provides better interference management. To demonstrate the effectiveness of RISs, the probability of outage of the considered network is investigated, and novel theoretical expressions are derived. 

Due to the nature of broadcasting, satellite networks are vulnerable to security threats similar to those of UAV and HAPS. In addition, assisting a satellite network with terrestrial/aerial RISs may increase the risk factor for these threats. Therefore, the authors of \cite{ActiveRISsecureGEO} develop a secure transmission scheme based on the PLS approach for an active terrestrial cognitive satellite network assisted by RIS. In the network, a satellite and a BS use the same spectrum resource, while multiple eavesdroppers try to capture the desired signal transmitted from the BS to a mobile user. Active RIS is exploited to improve the secure transmission of BS-user links and mitigate interference from the satellite. Maximization of the achievable secrecy rate is analyzed under the constraints of power and interference threshold. The authors have shown the domination of the RIS on artificial noise approach to achieve better security. In \cite{CovertRISsat}, covert communication is considered in a satellite network, where a terrestrial RIS improves the desired signals of legitimate users while hindering the illegitimate user. To maximize the minimum covert rate, a max-min optimization problem is formulated subject to transmit powers and the RIS's phase shifts. The authors of \cite{DoubleRISSecrecy} propose using a pair of RISs in uplink satellite networks, where a ground BS communicates with a legitimate satellite. To enhance the secrecy rate, one of the RISs is located near to the ground BS, while the other one is mounted on the legitimate satellite. An alternating optimization algorithm is proposed to jointly optimize the beamforming of the ground BS and phase shifts of the RISs. It is shown that the use of double RISs outperforms the case consisting of a single RIS in terms of secrecy rate.

Recently, LEO-based NTN have been regarded as the most notable solutions for enhancing 6G connectivity, because LEO satellites provide high throughput and low latency levels. However, terrestrial/aerial receivers experience significantly high Doppler shift effect because of the rapid movement of LEO satellites, even if those receivers are motionless. The Doppler shift brings about a mismatch of the carrier frequency offset (CFO) between the satellite and receiver, introducing increased inter-carrier interference (ICI) to the desired orthogonal frequency division multiplexing (OFDM) signal. Therefore, compensating for the Doppler effect is of utmost importance in LEO-based communication networks. Some methods used for terrestrial communication already exist and some of them are evaluated in \cite{DopplerOFDM}, such as \textit{ timing offset estimation and correction}, \textit{ frequency offset estimation and correction}, and \textit{ mitigation of ICI using frequency equalization / ICI self-canceling / time domain windowing}. For LEO-based satellite networks, the studies \cite{DopplerLeo1,DopplerLeo2} propose pre-compensating for the Doppler shift with respect to the LEO's beam center before the actual communication, while mitigating the residual Doppler shift portion at the receiver in a post-compensating manner. The methods used are shown to significantly improve the detection probability at the receiver; however, a high SNR level is required to estimate the residual Doppler shift in the high-frequency bands. Although the mentioned methods are able to partially mitigate the Doppler shift, they can increase signal overhead, hardware complexity, and transmission latency. However, utilizing RISs to compensate for the Doppler shift has been considered an alternative promising solution in recent literature \cite{BasarRISDoppler}. The authors of \cite{DopplerRISLEO1} propose deploying an RIS between a LEO satellite and the ground user for an uplink network to mitigate the Doppler shift effect by optimizing the phase shifts of the RIS. In \cite{JaeinLEO}, an optimization framework is proposed to improve the achievable rate in a network based on LEO aided by RIS. To mitigate the Doppler shift effect, power allocation optimization and passive beamforming for the RIS's phase shifts are jointly conducted. Likewise, the authors of \cite{TokaSPAWC} consider an OFDM-based LEO network and propose the deployment of a RIS between the satellite and the ground user. The outage probability of the system is analyzed together with asymptotic approximations to demonstrate the effectiveness of utilizing RISs for the mitigation of the Doppler shift effect. It has been shown that using RISs significantly enhances the received signal quality in the presence of the Doppler shift, if the phase shifts of RISs are properly adjusted.

\section{Conclusions and Future Prospects}


On one hand,  the \ac{NTN} could complement the terrestrial 6G infrastructure by extending coverage to remote and undeserved areas where deploying traditional \ac{TN} is challenging or economically unfeasible. On the other hand, increased propagation delays, Doppler shift, form factor, seamless coverage, etc., are some of the challenges faced by \ac{NTN} to become fully integrated in \ac{TN}. In the following, we have highlighted some of the key challenges faced by \ac{NTN} and some future considerations that can be taken into account to overcome them.

\subsection{Challenges}




\subsubsection{\textbf{Delay/Latency}} 
Due to satellite signals' considerable distance, they are always subject to high latency. To illustrate, round-trip times by \ac{GEO} satellites may exceed 500 milliseconds, which renders them unsuitable for those low or ultra-low latency applications including online gaming, video streaming, as well as virtual or augmented reality. In contrast, \ac{LEO} satellites have latencies that range from tens of milliseconds to tens of milliseconds. However, in case of the \ac{LEO}; realistic latency depends on a number of factors, such as satellite constellation size, which affects the waiting time for message sending, as well as sites, which influence payload delivery times around the ground. Compared to \ac{TN}, \ac{NTN} experience much higher propagation path losses due to the long distances to \ac{UE}. Additionally, \ac{NTN} platforms can cover large areas and serve a large number of users, leading to varied propagation delays and path losses in different regions. This variability makes it difficult to maintain consistent communication quality for all users and complicates the management of initial access and synchronization for this diverse user base \cite{gul2024ntn,varrall20185g}.


\subsubsection{\textbf{Channel Estimation}}
Channel estimation is a critical function in any communication system, but it presents unique challenges for \ac{NTN} due to their inherently time-varying characteristics. Spaceborne \ac{NTN} platforms, such as \ac{LEO} satellites, move across the sky from horizon to horizon within just 5-10 minutes. As a result, users on Earth are covered by a specific \ac{NTN} platform for only a short duration. Additionally, the long propagation delay can quickly make the estimated \ac{CSI} outdated. Consequently, conventional estimation methods used in \ac{TN} may not be appropriate for \ac{NTN}, requiring more advanced techniques to ensure efficient operation \cite{yuan2022secure,zhen2023clustering}.

\subsubsection{\textbf{Financial Viability/Device Cost}}
In the last few years, companies have made many announcements regarding the launch of \ac{LEO} satellite constellations. However, these launches often fall short of the big plans these firms set out, with many opting either not to proceed or to scale back their plans significantly. On one hand, the \ac{LEO} satellites faces a significant challenge with their short lifespan (ranges from 5 to 10 years). On the other hand, Companies deploying these satellites must rapidly generate enough business to keep up with the need to replace aging satellites and expand their constellations. Without swift business growth, companies may find it difficult to replace aging satellites and expand their constellations effectively \cite{kodheli2020satellite,pratt2019satellite}.






\subsubsection{\textbf{Resource management}}

\ac{NTN} platforms need to transmit signals with much higher power compared to terrestrial terminals to overcome significant path loss and ensure successful signal decoding for users on Earth \cite{khan2023resource}. This creates a challenge, since the \ac{NTN} platforms do not have access to the stable power sources available to the terrestrial terminals. Furthermore, the frequency bands designated for \ac{NTN} communications, namely the S-band and the Ka-band, are limited and already heavily used. The S band is occupied by 4G \ac{LTE} devices, while the Ka-band is used by devices equipped with millimeter wave in 5G. As a result, \ac{NTN} users may face co-channel interference from these terrestrial devices. This situation calls for innovative spectrum sharing solutions to intelligently and efficiently manage the limited frequency bands \cite{kodheli2020satellite,pratt2019satellite}.

\subsubsection{\textbf{Mobility Management/Doppler Effect}}

Users on Earth face major Doppler effects from \ac{NTN} platforms due to their fast movement speeds, which cause significant distortion in communications links. The Doppler effect, which involves a frequency shift of signals caused by relative motion between transceivers, also occurs in \ac{TN}, such as with users in high-speed trains or cars. However, this effect is more pronounced in \ac{NTN} due to the higher velocities of these platforms \cite{larrayoz2024ml}. For example, a user communicating with a \ac{LEO} satellite at an altitude of 600 kilometers may experience a Doppler shift of up to 48 kHz at a carrier frequency of 2 GHz, which is much larger than the shifts typically encountered in \ac{TN} \cite{kodheli2020satellite}.

\subsection{Future Considerations}

\subsubsection{\textbf{Integrated NTN \& TN}}
6G is anticipated to deliver significantly enhanced connectivity compared to previous generations and offer better data rates, reduced latency and stronger reliability. In order to realize this ambition \ac{NTN} will be merged with \ac{TN} so that seamless connectivity will be maintained across different regions. One of the intrinsic disadvantages of \ac{NTN} includes limited uninterrupted coverage within some places like buildings, heavily forested areas or highly populated cities. The combination of \ac{NTN} and \ac{TN} helps solve the problem of ensuring continuous coverage. Creating hybrid solutions that utilize both cellular phone and satellite technology has an advantage over merely relying on satellite communication \cite{shang2024multi,msadaa2022non}.

\subsubsection{\textbf{Orthogonal Time Frequency Space Modulation for NTN}}

To guarantee establishing high-reliability communication in NTN deployments, novel methods having the ability to deal with the Doppler effect are required. Although the OFDM scheme is currently used in many mobile networks; however, it is sensitive to ICI caused by the Doppler effect. In addition, the limited coherence time of the time-frequency domain channel in an OFDM system causes a significant channel estimation overhead. It is shown that one channel coherence interval may contain only a few OFDM symbols, causing an increase in the number of pilot symbols \cite{OTFS1}. To address this issue, orthogonal time frequency space (OTFS) modulation has recently been proposed as an efficient solution. Unlike the OFDM scheme, the OTFS modulation is a two-dimensional scheme considering the delay-Doppler domain to multiplex information signals. Hence, OTFS modulation offers considerable benefits, such as delay / Doppler resilience, robustness against CFO (introducing ICI), alleviated channel estimation overhead, enhanced spectrum utilization, reduced pick-to-average power ratio, and flexible interaction with existing systems, which encourage researchers to exploit it in high-mobility systems \cite{OTFS2,OTFS3}.

\subsubsection{\textbf{Blockchain for NTN}}
Blockchain technology offers considerable promise for improving multiple facets of \ac{NTN} systems, including security, interoperability, efficiency, and privacy \cite{jameel2020minimizing}. For example, utilizing public and private key mechanisms along with digital signatures, the blockchain can enhance secure and privacy-oriented identity management within \ac{NTN}. This approach facilitates transparent administration and protects against unauthorized access. In addition, smart contracts can streamline network segmenting by automating agreements between network entities and optimizing the allocation and management of resources. The blockchain also promotes collaboration and resource sharing through digital tokens and incentive structures, motivating participants to contribute resources, share data, or collaborate, thus addressing the challenges related to resource and data management in \ac{NTN}.

\subsubsection{\textbf{Generative AI for NTN}}
Exploring the integration of Generative \ac{AI} into \ac{NTN} is a promising research direction with substantial potential. Generative \ac{AI}, known for its ability to create synthetic data and novel content, can address several challenges faced by \ac{NTN}. For example, it can produce high-quality synthetic data to support the training and testing of traditional \ac{AI} methods in the complex and diverse environments of \ac{NTN}. Additionally, Generative \ac{AI} can enhance \ac{NTN} security by generating realistic attack scenarios, which aids in evaluating the network defense mechanisms thoroughly. Moreover, it can be used to generate network data to simulate various conditions, thereby facilitating network optimization and predicting potential bottlenecks \cite{jamshed2024non}.

\subsection{Conclusion}

The introduction of \ac{NTN} technology entails numerous challenges; however, it is important to note that these difficulties are far outweighed by the benefits. Successfully implementing \ac{NTN} requires a coordination of government action, private sector initiatives, and international organizations to set up the infrastructure and regulatory frameworks required. Some of the major benefits that \ac{NTN} can bring include improved human development, economic advancement, and global connectivity. In essence, a powerful solution to connectivity challenges is what offers \ac{NTN}. Its potential to connect underserved populations via extensive resilient cell phone coverage, enhance disaster mitigation ability, and support worldwide \ac{IoT} applications makes it capable of closing the digital gap. The fact that this technology has the ability to empower marginalized groups, predominantly women from developing countries, emphasizes its revolutionary role in rural areas. This could be a critical factor in ensuring that nobody is left behind in terms of accessing digitally connected life as the world moves towards greater interdependence between nations.

\section*{Acknowledgements}
M. Z. Shakir and S. Manzoor acknowledge that this publication was made possible by NPRP13S-0130-200200 from the Qatar National Research Fund (a member of the Qatar Foundation).  The work of M. Toka, J. Seong, and W. Shin was supported by the Institute of Information \& Communications Technology Planning \& Evaluation (IITP) grants (No.2021-0-00260 and No.2021-0-00467).


\bibliographystyle{IEEEtran}

\bibliography{IEEEabrv,BibRef}


\end{document}